\documentclass[12pt]{article}

\usepackage{amsmath,amssymb,amsthm,amscd,pstricks}
\usepackage{graphicx,tocloft}
\usepackage[nosort]{cite}
\def\figurename{Figure}

\makeatletter

\renewcommand{\fnum@figure}[1]{\textbf{\figurename~\thefigure}:}

\makeatother

\makeatletter

\renewcommand\section{\@startsection{section}{1}{\z@}
                                   {-3.5ex \@plus -1ex \@minus -.2ex}
                                   {2.3ex \@plus .2ex}
                                   {\normalfont\large\bfseries}}
\renewcommand\subsection{\@startsection{subsection}{2}{\z@} 
                                   {-3.25ex\@plus -1ex \@minus -.2ex}
                                   {1.5ex \@plus .2ex}
                                   {\normalfont\normalsize\bfseries}}
\renewcommand\subsubsection{\@startsection{subsubsection}{3}{\z@}
                                   {-3.25ex\@plus -1ex \@minus -.2ex}
                                   {1.5ex \@plus .2ex}
                                   {\normalfont\normalsize\bfseries}}
\renewcommand\paragraph{\@startsection{paragraph}{4}{\z@}
                                   {3.25ex \@plus1ex \@minus.2ex}
                                   {-1em}
                                   {\normalfont\normalsize\bfseries}}

\makeatother

\newdimen\tableauside\tableauside=1.0ex
\newdimen\tableaurule\tableaurule=0.4pt
\newdimen\tableaustep
\def\phantomhrule#1{\haox{\vbox to0pt{\hrule height\tableaurule
width#1\vss}}}
\def\phantomvrule#1{\vbox{\haox to0pt{\vrule width\tableaurule
height#1\hss}}}
\def\sqr{\vbox{%
  \phantomhrule\tableaustep

\haox{\phantomvrule\tableaustep\kern\tableaustep\phantomvrule\tableaustep}%
  \haox{\vbox{\phantomhrule\tableauside}\kern-\tableaurule}}}
\def\squares#1{\haox{\count0=#1\noindent\loop\sqr
  \advance\count0 by-1 \ifnum\count0>0\repeat}}
\def\tableau#1{\vcenter{\offinterlineskip
  \tableaustep=\tableauside\advance\tableaustep by-\tableaurule
  \kern\normallineskip\haox
    {\kern\normallineskip\vbox
      {\gettableau#1 0 }%
     \kern\normallineskip\kern\tableaurule}%
  \kern\normallineskip\kern\tableaurule}}
\def\gettableau#1 {\ifnum#1=0\let\next=\null\else
  \squares{#1}\let\next=\gettableau\fi\next}

\newcommand{\be}{\begin{equation}}
\newcommand{\ee}{\end{equation}}
\newcommand{\bea}{\begin{eqnarray}}
\newcommand{\eea}{\end{eqnarray}}
\newcommand{\ba}{\begin{array}}
\newcommand{\ea}{\end{array}}
\newcommand{\id}{\haox{1\kern-.27em l}}
\newcommand{\ZZ}{\mathbb{Z}}
\newcommand{\CC}{\mathbb{C}}
\newcommand{\RR}{\mathbb{R}}
\newcommand{\half}{ {\textstyle \frac{1}{2}  } }
\newcommand{\al}{\alpha}

\newcommand{\ka}{\kappa}

\newcommand{\de}{\delta}

\newcommand{\ep}{\epsilon}

\newcommand{\si}{\sigma}
\newcommand{\la}{\lambda}

\newcommand{\De}{\Delta}
\newcommand{\La}{\Lambda}

\newcommand{\Ups}{\Upsilon}
\newcommand{\cN}{\mathcal{N}}

\newcommand{\cQ}{\mathcal{Q}}
\newcommand{\cK}{\mathcal{K}}
\newcommand{\cW}{\mathcal{W}}

\newcommand{\cV}{\mathcal{V}}

\newcommand{\tr}{{\rm tr}}

\newcommand{\pa}{\partial}
\newcommand{\rar}{\rightarrow}

\newcommand{\non}{\nonumber}
\newcommand{\lb}{\langle}
\newcommand{\rb}{\rangle}
\newcommand{\re}{{\rm e}}

\newcommand{\tia}{\tilde{a}}
\newcommand{\tij}{\tilde{\jmath}}
\newcommand{\SU}{\mathrm{SU}}

\newcommand{\SL}{\mathrm{SL}}

\newcommand{\gl}{\mathrm{gl}}
\newcommand{\sll}{\mathrm{sl}}
\newcommand{\U}{\mathrm{U}}

\newcommand{\ts}{\textstyle}

\begin{document}
\begin{flushright} {CERN-PH-TH/2010-177} \\ HU-EP-10/47\end{flushright}

\begin{center}
\vspace*{4mm}
{\Large\sf
{Affine $\sll${\large$(N)$} conformal blocks from {\large $\cN=2$} {\large $\SU(N)$} gauge theories 
}}

\vspace*{7mm}
{\large Can Koz\c{c}az${}^\dagger$, Sara Pasquetti${}^\dagger$, Filippo Passerini$^\S$ and Niclas Wyllard}

\vspace*{5mm}
       ${}^\dagger$ PH-TH division, CERN, CH-1211 Geneva, Switzerland \\
\vspace*{3mm}
       ${}^\S$  Institut f\"ur Physik, Humboldt-Universit\"at zu Berlin,\\
Newtonstra{\ss}e 15, D-12489 Berlin, Germany

\vspace*{5mm}

{\tt Can.Kozcaz@cern.ch, Sara.Pasquetti@cern.ch,  \\	
filippo@physik.hu-berlin.de, n.wyllard@gmail.com}

\vspace*{12mm}
{\bf Abstract} 
\end{center}
\vspace*{0mm}
\noindent  Recently Alday and Tachikawa \cite{Alday:2010} proposed a relation between conformal blocks in a two-dimensional theory with affine $\sll(2)$ symmetry and instanton partition functions in four-dimensional conformal $\cN=2$ $\SU(2)$ quiver gauge theories in the presence of a certain surface operator. In this paper we extend this proposal  to a relation between conformal blocks in theories with affine $\sll(N)$ symmetry and instanton partition functions in conformal $\cN=2$ $\SU(N)$ quiver gauge theories in the presence of a surface operator. We also discuss the extension to non-conformal $\cN=2$ $\SU(N)$ theories.

\vspace{1mm}

\setcounter{tocdepth}{1}
\tableofcontents

\setcounter{equation}{0}
\section{Introduction}\label{sint}

Ever since the groundbreaking work of Seiberg and Witten \cite{Seiberg:1994a}, the study of four-dimensional gauge theories with $\cN=2$ supersymmetry has been an important research topic. Such theories have a very rich structure and have many remarkable connections to other areas of both physics and mathematics. 

Last year, building on earlier work by Witten \cite{Witten:1995}, Gaiotto \cite{Gaiotto:2009a} introduced a new way of analysing $\cN=2$ theories by viewing them as arising from a six-dimensional theory compactified on a two-dimensional Riemann surface with punctures. In this approach one naturally expects connections between the $4d$ $\cN=2$ gauge theory and some conformal theory on the $2d$ Riemann surface. 

The AGT relation  \cite{Alday:2009a} is a precise realisation of this expectation. It encompasses a relation between instanton partition functions in conformal $\cN=2$ theories and conformal blocks in two-dimensional conformal field theories.  The original work \cite{Alday:2009a} proposed a relation between conformal $4d$ $\SU(2)$ quiver gauge theories and the $2d$ Liouville theory. This relation was subsequently extended  \cite{Wyllard:2009} to a relation between conformal $4d$ $\cN=2$ $\SU(N)$ theories and $2d$ $A_{N-1}$ Toda field theories. Non-conformal $\cN=2$ theories have also been considered and related to two-dimensional CFT \cite{Gaiotto:2009b,Marshakov:2009}. 
  
 A natural way to extend the AGT relation is to consider the inclusion of various defects in the gauge theory.  Examples include one-dimensional (line) defects (e.g.~Wilson and 't Hooft loops), and three-dimensional (domain wall) defects. Such defects have been considered in  \cite{Alday:2009b,Drukker:2009}, and \cite{Drukker:2010}, respectively. 

In this paper we focus on defects which are supported on two-dimensional submanifolds, i.e.~surface operators. Surface operators in $\cN=4$ gauge theories were extensively studied in \cite{Gukov:2006} (see  \cite{Gukov:2007} for some similar work in $\cN=2$ theories). In the context of the AGT relation, surface operators  have been studied in several papers \cite{Alday:2009b,Gaiotto:2009c,Kozcaz:2010,Dimofte:2010,Maruyoshi:2010,Taki:2010,Alday:2010}. 

When viewed from the six-dimensional perspective there are two ways a surface operator can arise \cite{Alday:2010}: either from a $4d$ defect wrapping the $2d$ Riemann surface, or as a $2d$ defect intersecting the $2d$ Riemann surface at a point. The second class of surface operators can be described in the dual $2d$ CFT  by inserting a certain degenerate field operator localised at a point. Such surface operators were first considered in \cite{Alday:2009b} and have been further studied in \cite{Gaiotto:2009c,Kozcaz:2010,Dimofte:2010,Maruyoshi:2010,Taki:2010}. 
For the first class of surface operators it was recently  proposed \cite{Alday:2010} that the effect of wrapping the $4d$ defect around the Riemann surface is to modify the $2d$ CFT to another  $2d$ CFT. For the $\SU(2)$  quiver gauge theories it was argued that the surface operator insertion modifies the dual Liouville theory  to a theory with (untwisted) affine $\sll(2)$ symmetry.

Conformal blocks in this theory should therefore be related to 
 instanton partition functions in $\SU(2)$  quiver gauge theories in the presence of a surface operator  \cite{Alday:2010} that arises from a $4d$ defect.

 It was further realised in \cite{Alday:2010} that the technology to compute such instanton partition functions already exists in the mathematics literature \cite{Braverman:2004a,Negut:2008,Feigin:2008}. Using these results several checks of the proposed relation were performed. 

In this paper we extend the proposal in \cite{Alday:2010} to a relation between conformal blocks in theories with affine $\sll(N)$ symmetry and instanton partition functions in conformal $\cN=2$ $\SU(N)$ quiver gauge theories in the presence of a 
surface operator arising from a $4d$ defect. In other words, we argue that the effect of the $4d$ defect is to replace the $A_{N-1}$ Toda field  theory and its associated $\cW_{N}$-algebra symmetry by a theory with affine $\sll(N)$ symmetry. We perform several checks of the proposed relation and also extend it to non-conformal $\cN=2$ $\SU(N)$ theories. 

 In the next section we review some facts about instanton counting in $\SU(N)$ quiver gauge theories in the presence of a surface operator, and in section \ref{ssl2} we review the proposal in \cite{Alday:2010} and perform some additional tests using a different perturbative scheme compared to the one in \cite{Alday:2010} which allows us to sum up certain infinite sets of terms. For the rank one case we also discuss the relation to the surface operator arising from  a degenerate field insertion in the Liouville theory.  Then in section \ref{sslN} we propose a relation between conformal blocks in a theory with affine $\sll(N)$ symmetry and instanton partition functions in $\SU(N)$ quiver gauge theories with a surface operator insertion. The extension to non-conformal theories is discussed in section \ref{snas}. In the appendix some technical details are collected.

{\bf Note added:} After this work was finished \cite{Awata:2010} appeared. This paper has some overlap with our results, but only considers the case of $\SU(2)$.

\setcounter{equation}{0}
\section{Surface operators and instanton counting}\label{sinst}

A surface operator in a four-dimensional gauge theory is a certain object supported on a two-dimensional submanifold of spacetime.  One way to define a surface operator is by specifying the (singular) behaviour of the gauge field (and scalars, if present) near the submanifold where the surface operator is supported. An extensive study of surface operators in the context of the $\cN=4$ $\SU(N)$ gauge theories (in a flat spacetime) was carried out in \cite{Gukov:2006}. There it was found that the possible types of surface operators supported on an $\RR^2$ submanifold  are in one-to-one correspondence with the so called Levi subgroups (whose classificiation in turn is in one-to-one correspondence with the various (non-trivial) ways of embedding $\SU(2)$ inside $\SU(N)$, or equivalently the number of possible ways of breaking $\SU(N)$ to a $\U(1)^{\ell-1} \prod_{i=1}^{\ell} \SU(N_i)$ (proper) subgroup). Concretely this means that for every (non-trivial) partition $N=N_1+\ldots+N_\ell$ there is a possible surface operator. In this paper we study surface operators\footnote{Throughout we assume that the surface operator is supported on an $\RR^2$ submanifold.} in $4d$ $\SU(N)$ theories with $\cN=2$ supersymmetry;  such surface operators are also classified by the Levi subgroups.  For $\cN=2$ theories a surface operator depends on  a certain number of continuous complex parameters, one for each of the abelian $\U(1)$ factors in the Levi subgroup (unbroken group)\footnote{For $\cN=4$ theories the surface operators depend on four real parameters for each $\U(1)$ factor.}.

In \cite{Alday:2010} the following terminology was used: a {\it full} surface operator corresponds to the  breaking of $\SU(N)$ to  $\U(1)^{N-1}$ and depends on $N-1$ continuous parameters (this is the maximal number of parameters possible), whereas a {\it simple} surface operator corresponds to the  breaking of  $\SU(N)$ to $\SU(N{-}1){\times}\U(1)$ and depends on one parameter.

A surface operator with a given Levi type of singularity can  be realised
both by $4d$ or by $2d$ defects, in the $6d$ language.
In particular there will be  {\it full} surface operators coming from $2d$ and $4d$ defects as well as 
 {\it simple} surface operators coming from $2d$ and $4d$ defects.
Different realisations are not supposed to give rise to the same surface operator, however one may speculate that
the instanton partition function may not be sensitive to the difference.
We will explore this possibility in section \ref{m2m5}.

 In this paper  surface operators that arise (in the $6d$ language) from $4d$ defects will always be full surface operators, whereas the surface operators that arise from $2d$ defects are simple surface operators. Sometimes, for convenience  we will refer to  the two classes just as full and simple surface operators, respectively. 
But the reader should keep in mind that there are in general two realisations for each Levi type of singularity.

\subsection{$\SU(N)$ instanton counting in the presence of a simple surface operator}

A natural question to address is how the instanton partition function in an $\cN=2$ gauge theory \cite{Nekrasov:2002} (which is valid in the absence of surface operators) changes when a surface operator is present. 

In \cite{Alday:2009b} it was conjectured that a simple surface operator in a (mass-deformed) conformal $\SU(2)$ theory  has a dual description in the Liouville theory in terms of  the insertion of a certain degenerate field.  It was shown that in a semi-classical limit this implies that the effect of the simple surface operator in the gauge theory can be computed from the Seiberg-Witten data, i.e.~the curve and the differential.  
In a further development \cite{Kozcaz:2010} it was shown how to go beyond the semi-classical analysis performed in \cite{Alday:2009b} in an order-by-order (``B-model") expansion (this method  also works for the cases where several simple surface operators are present). 

In \cite{Kozcaz:2010} it was also shown  that by combining the conjectures in \cite{Alday:2009a} and \cite{Alday:2009b} (using also a result in \cite{Mironov:2009a}) one can obtain (conjectural) closed expressions for the gauge theory instanton partition function in $\SU(N)$ theories when simple surface operators are present (this method also works for the non-conformal cases). When lifted to $5d$ these instanton partition functions have a natural (``A-model")  topological string interpretation. As emphasized by Gukov, in the topological string language a simple surface operator corresponds to a  toric brane. Computing topological string partition functions with toric brane insertions leads to agreement \cite{Kozcaz:2010,Dimofte:2010} with what one obtains from the combination of the conjectures in \cite{Alday:2009a} and \cite{Alday:2009b}.
In particular, in \cite{Dimofte:2010} it was argued that in the topological string language this type of conjectured duality corresponds to a geometric transition (see also \cite{Taki:2010}).

For an arbitrary surface operator, generic features of the instanton expansion were discussed in \cite{Alday:2009b}. For a full surface operator one can obtain exact results as we discuss next.

\subsection{$\SU(N)$ instanton counting in the presence of a full surface operator}

In a recent  paper \cite{Alday:2010} Alday and Tachikawa proposed  that the formalism needed to  determine the instanton partition function in the presence of a full surface operator in an $\SU(N)$ theory has already been developed in the mathematical literature \cite{Braverman:2004a,Negut:2008,Feigin:2008}. (Strictly speaking, it is not completely obvious that the problem solved by the mathematicians is really equivalent to the physics problem, but this is believed to be the case.) 

Before we describe this construction it is convenient to first briefly recapitulate some relevant facts about the partition function, $Z$, in an $\cN=2$ $\SU(N)$ quiver gauge theory (without surface operators). The partition function contains all information about the low-energy effective action and contains both perturbative (classical and one loop levels only) as well as instanton contributions; in other words
\be
Z = Z_{\rm pert}\,Z_{\rm inst} \,.
\ee
The Nekrasov instanton partition function $Z_{\rm inst}$ is obtained from certain (regularised) integrals   over the moduli space of instantons (first studied in \cite{Moore:1997}). The regularisation involves two deformation parameters, $\ep_1$ and $\ep_2$, that ensure that these integrals localise to isolated fixed points and can be explicitly evaluated in closed form \cite{Nekrasov:2002}.  The fixed points are labelled by a vector of Young tableaux, $\la=(\la^1,\ldots,\la^N)$  \cite{Nekrasov:2002}, and the resulting instanton partition function takes the form
\be
Z_{\rm inst}= \sum_{\la}  Z_{k}(\la) \, y^{k} \,,
\ee
where the sum is over all vectors of Young tableaux $\la$, and  
the instanton number $k= |\la|$ is equal to the sum of the boxes in all the $\la^i$. 

In general, a succinct way to summarise the result is in terms of a certain character. The character encodes the contribution to the instanton partition function from a given fixed point and takes the general form
\be \label{char}
\chi = \sum_i (\pm) e^{w_i} \,.
\ee
The contribution to the instanton partition function from the given fixed point  (denoted $Z_{k}(\la)$ above) is given by the product over the weights $w_i$ where  the weights coming from terms in (\ref{char}) with a minus sign contribute in the denominator and those arising from terms with a plus sign contribute in the numerator.

A basic building block is the character for a hypermultiplet  of mass $m$ transforming in the bifundamental  representation of $\SU(N){\times}\SU(N)$, which is of the general form
\be \label{bifchar}
\chi_{\rm bif} (a,\tia,\la,\xi,m) \,.
\ee
(The precise form can be found in \cite{Fucito:2004}, but will not be needed in this paper.)  
In the expression (\ref{bifchar}), $a=(a_1,\ldots,a_N)$ are the Coulomb moduli of the first $\SU(N)$ factor in the gauge group and  $\la=(\la^1,\ldots,\la^N)$ is a vector of Young tableaux referring to the same $\SU(N)$ factor;  $\xi=(\xi^1,\ldots,\xi^N)$ is a vector of Young tableaux referring to the second $\SU(N)$ factor and $\tia=(\tia_1,\ldots,\tia_N)$ are the associated Coulomb moduli.  Since we want the gauge group to be $\SU(N)$ we need to impose (by hand) the restriction $\sum_i a_i =0$ (and similarly for the $\tia_i$'s). 

From the expression (\ref{bifchar}) one can obtain the character for other representations of interest such as the character for $N$ hypermultiplets transforming in the fundamental representation of the first (or second) $\SU(N)$ factor, which are arise from  
\bea \label{Nfunds}
\chi_{N \, {\rm funds}}(a,\la,\tilde{\mu}) = \chi_{\rm bif} (a,\tilde{\mu},\la,\emptyset,0) \,,\non \\
\chi_{N \, {\rm funds}}(\tia,\xi,\mu) = \chi_{\rm bif} (\mu,\tia,\emptyset,\xi,0)   \,,
\eea
where $\mu=(\mu_1,\ldots,\mu_N)$ and $\tilde{\mu}=(\tilde{\mu}_1,\ldots,\tilde{\mu}_N)$ denote the masses of the fundamentals {\it without} any restriction on $\sum_i \mu_i$ and $\sum_i \tilde{\mu}_i$, and transform under a $\U(N)$ flavour symmetry. (Alternatively, one can decompose $\mu$ into a part transforming under an $\SU(N)$ flavour symmetry plus an additional mass parameter transforming under a $\U(1)$ flavour symmetry.)

The character for a matter multiplet of mass $m$ transforming in the adjoint representation of $\SU(N)$ is given by
\be \label{adj}
\chi_{\rm adj} (a,\la,m) = \chi_{\rm bif} (a,a,\la,\la,m)  \,,
\ee
and finally the character of the gauge vector multiplet of $\SU(N)$ is obtained via
\be \label{vec}
\chi_{\rm vec} (a,\la) = -\chi_{\rm bif} (a,a,\la,\la,0)  \,.
\ee

Just as in the absence of surface operators, the instanton partition function in an $\SU(N)$ theory  with a full surface operator involves a sum over a certain $N$-dimensional vector of Young tableaux $\la=(\la^1,\ldots,\la^N)$ where each $\la^i$ denotes a Young tableau, or equivalently, a partition\footnote{In contrast to \cite{Feigin:2008} we label the components, $\la^i_j,$  of $\la^i$ starting from $j=1$ rather than $j= 0$.}, i.e.~$\la^i_1 \ge \la^i_2 \cdots$.

 It turns out to be very convenient to view the partitions as having a periodicity, $\la^i \equiv \la^{i+N}$. Similarly, the Coulomb moduli are assumed to have the same property: $a_i \equiv a_{i+N}$. 
 The character for a bifundamental multiplet can then be written \cite{Feigin:2008,Alday:2010}   
\bea \label{surfchar}
&&  \!\!  \!\!  \!\!  \!\!  \!\!  \chi_{\rm bif} (a,\tia,\la,\xi,m) \,  = \, e^{-m}\sum_{k=1}^{N} \sum_{\ell' \ge 1} e^{a_{k}-\tia_{k-\ell'}}e^{\ep_2( \lfloor \frac{\ell'-k}{N} \rfloor -  \lfloor -\frac{k}{N} \rfloor)}  \!\!\!\!\!  \sum_{s=1}^{\quad \xi^{k-\ell'}_{\ell'}}  \!\!\!  e^{\ep_1s } \non \\
&-&\!\!  e^{-m}\sum_{k=1}^{N} \sum_{\ell\ge 1}\sum_{\ell' \ge 1} e^{a_{k-\ell+1}-\tia_{k-\ell'}}e^{\ep_2( \lfloor \frac{\ell'-k}{N} \rfloor -  \lfloor \frac{\ell-k-1}{N} \rfloor)} (e^{\ep_1\xi^{k-\ell'}_{\ell'}}-1) \!\!\!\!\!  \sum_{s=1}^{\quad \la^{k-\ell+1}_{\ell}}  \!\!\!  e^{\ep_1(s-\la^{k-\ell+1}_{\ell})} \non \\
&+&\! \! e^{-m}\sum_{k=1}^{N} \sum_{\ell\ge 1}\sum_{\ell' \ge 1} e^{a_{k-\ell+1}-\tia_{k-\ell'+1}}e^{\ep_2( \lfloor \frac{\ell'-k-1}{N} \rfloor -  \lfloor \frac{\ell-k-1}{N} \rfloor)} (e^{\ep_1\xi^{k-\ell'+1}_{\ell'}}   \!\!\!\ -1) \!\!\!\!\!  \sum_{s=1}^{\quad \la^{k-\ell+1}_{\ell}}  \!\!\!  e^{\ep_1(s-\la^{k-\ell+1}_{\ell})}  \non  \\
&+&\!\!  e^{-m}\sum_{k=1}^{N} \sum_{\ell\ge 1} e^{a_{k-\ell+1}-\tia_{k}}e^{\ep_2( \lfloor -\frac{k}{N} \rfloor -  \lfloor \frac{\ell-k-1}{N} \rfloor)} \!\!\!\!\!  \sum_{s=1}^{\quad \la^{k-\ell+1}_{\ell}}  \!\!\!  e^{\ep_1(s-\la^{k-\ell+1}_{\ell})} 
\eea
where $\lfloor x \rfloor$ denotes the largest integer smaller than or equal to $x$. 

From the result (\ref{surfchar}) one can obtain the character for $N$  hypermultiplets transforming in the fundamental representation of the first gauge group by setting $\xi^j=\emptyset$ for all $j$, cf.~(\ref{Nfunds}). Similarly, for $N$ hypers in the fundamental representation of the second factor one sets $\la^i=\emptyset$, cf.~(\ref{Nfunds}). (The masses of the fundamentals are assumed to have the same periodicity as the Coulomb moduli and the partitions, i.e.~$\mu_i=\mu_{i+N}$ etc.) The character for a massive matter multiplet transforming in the adjoint can also easily be obtained, cf.~(\ref{adj}). Finally, the contribution from a gauge vector multiplet is obtained by setting $\xi=\la$ and $m=0$, cf.~(\ref{vec}).

From these building blocks the instanton partition function for an $\SU(N)$ quiver gauge theory with bifundamental and fundamental matter multiplets in the presence of a full surface operator can be determined. For a  gauge group with a single $\SU(N)$ factor the result is of the form
\be
Z_{\rm inst}= \sum_{\la}  Z_{k_1,\ldots,k_{N}}(\la) \prod_i y_i^{k_i} \,,
\ee
where the instanton numbers $k_i$ are given by \cite{Feigin:2008,Alday:2010}
\be \label{ki}
k_i = \sum_{j \ge 1} \la^{i-j+1}_j \,,
\ee
and the variables $y_i$ (defined for $i=1,\ldots,N$ and not assumed to be periodic in $i$) correspond to the $N-1$ (holomorphic) parameters of the full surface operator together with the usual instanton expansion parameter. 
In the general case of a quiver gauge group with several $\SU$ factors, there is a set of $y_i$ and $k_i$ for each factor, thus a full surface operator corresponds to breaking the complete gauge group to $\U(1)^r$ where $r$ is the sum of the ranks of all factors of the quiver gauge group. 

Next we consider in more detail three examples with a single $\SU(N)$ factor: the pure $\SU(N)$ theory, as well as two superconformal theories, the $\cN=2^*$ theory (i.e.~the theory with an adjoint matter multiplet), and the theory with $N_f=2N$ (i.e.~$2N$ matter multiplets in the fundamental representation). 

First we consider the terms with only one $k_i$ non-zero. In this case, one easily sees from (\ref{ki}) that only  $\la^i$ can be non-zero and furthermore can have boxes only in the first column, 
i.e.~only $\la^i_1$ is $\neq 0$ . This is because a non-zero $\la^j$ with $j\neq i$ inevitably makes at least one $k_j$ with $j\neq i$ non-zero, and the same is true for a non-zero $\la^i_j$ with $j\ge 2$.  With only $\la^i$ non-zero and composed of only one column of height $n\equiv \la^i_1$, there is only one contribution at each order in the instanton expansion. From (\ref{surfchar}) we find that for the  $\cN=2^*$ $\SU(N)$ theory the character corresponding to the $y_i^n$ term in the instanton expansion becomes
\bea
&& (e^{-m}-1)(e^{a_{i+1}-a_i} +1)\sum_{s=1}^{n} e^{\ep_1 s}  \quad \qquad (i\le N-1) \non \\
&& (e^{-m}-1)(e^{a_{i+1}-a_{i} + \ep_2} +1) \sum_{s=1}^{n} e^{\ep_1 s}  \qquad (i = N) 
\eea
(for the pure $\SU(N)$ theory the result is the same but the terms involving $e^{-m}$ are absent), whereas for the $\SU(N)$ theory with $N_f=2N$ one finds 
\bea
&&(-e^{a_{i+1}-a_i} +e^{\mu_{i+1}-a_i} + e^{a_{i}-\tilde{\mu}_i-\ep_1 n} -1)\sum_{s=1}^{n} e^{\ep_1 s}  \qquad \qquad \;  (i\le N-1) \non \\
&& (-e^{a_{i+1}-a_i+ \ep_2} +e^{\mu_{i+1}-a_i+ \ep_2} + e^{a_{i}-\tilde{\mu}_i-\ep_1 n} -1) \sum_{n=1}^{s} e^{\ep_1 s}  \qquad (i = N) 
\eea
These results lead to the following terms in the instanton partition function  for the pure $\SU(N)$ theory
\be \label{Z0pure}
Z_{\rm inst}^{(0,i)} = \sum_{n=1}^{\infty} \frac{1}{(\frac{a_{i+1}}{\ep_1}-\frac{a_{i}}{\ep_1}+\frac{\ep_2}{\ep_1}\lfloor \frac{i}{N}\rfloor+1)_n \, n!}  \left( \! \frac{y_i}{(\ep_1)^2} \! \right)^n .
\ee
Similarly, for the $\cN=2^*$ $\SU(N)$ theory one gets
\be \label{N2*Z0}
Z_{\rm inst}^{(0,i)} = \sum_{n=1}^{\infty} \frac{(\frac{a_{i+1}}{\ep_1}-\frac{a_{i}}{\ep_1}+\frac{\ep_2}{\ep_1}\lfloor \frac{i}{N}\rfloor+1-\frac{m}{\ep_1})_n (1-\frac{m}{\ep_1})_n}{(\frac{a_{i+1}}{\ep_1}-\frac{a_{i}}{\ep_1}+\frac{\ep_2}{\ep_1}\lfloor \frac{i}{N}\rfloor+1)_n \, n!} \, (y_i)^n \,,
\ee
whereas for the $\SU(N)$ theory with $N_f=2N$ the result is
\be \label{Nf4Z0}
Z_{\rm inst}^{(0,i)} = \sum_{n=1}^{\infty} \frac{(\frac{\mu_{i+1}}{\ep_1}-\frac{a_{i}}{\ep_1}+\frac{\ep_2}{\ep_1}\lfloor \frac{i}{N}\rfloor + 1)_n(\frac{\tilde{\mu}_i}{\ep_1}-\frac{a_{i}}{\ep_1})_n}{(\frac{a_{i+1}}{\ep_1}-\frac{a_{i}}{\ep_1}+\frac{\ep_2}{\ep_1}\lfloor \frac{i}{N}\rfloor+1)_n \, n!} \, (-y_i)^n  \,.
\ee
In the latter two cases, $Z_{\rm inst}^{(0,i)}$ is a hypergeometric function of the form ${}_2F_1(A ,B ; C ;y_i)$. 

It is also possible to write down corrections to the above results. One natural class of corrections involve terms of the form $y_i^n \, y_{j}$ with $i\neq j$. Terms of this type get contributions from at most two types of Young tableaux at each order. One always gets a contribution when  $\la^i$ has only one column with $n$ boxes and $\la^{j}$ contains only one box, with all other $\la^k$ empty. In addition, there are two special cases. First, when $j=i+1$ one gets a contribution when $\la^i$ has $n$ boxes in the first column and one box in the second column with all other $\la^k$ empty. Second, when $i=j+1$ one gets a contribution when $\la^i$ has $n-1$ boxes in the first column  and $\la^{i-1}$ has one box in both the first and second columns, with all other $\la^k$ empty.
As the resulting formul\ae{} are somewhat lengthy they have been relegated to the appendix, cf.~(\ref{Z1pure}), (\ref{nf41}).

Because of the presence of $\lfloor\cdot\rfloor$ in the above formul\ae{}, the terms involving $y_N$ are treated differently compared to the terms involving only the other $y_i$. We will see in later sections that this result is reflected in the affine conformal blocks where the worldsheet coordinate  $z$ is on a different footing compared to the isospin $x_i$ variables. The terms in the instanton partition function that are independent of $y_N$ form an important subsector that was studied in \cite{Negut:2008}. Such terms  have $k_N=0$, which by (\ref{ki}) implies that $\la_{N-j+1}^{j}=0$. Thus only a finite number of components of each $\la^j$ can be non-zero. In this case the character (\ref{surfchar}) can be simplified. One finds after some algebra that
\bea
\chi_{\rm bif} (a,\tia,\la,\xi,m)\left|_{k_N=0} \right. \! &=& \!  e^{-m}\sum_{k=1}^{N-1}\sum_{j=1}^{k+1}\sum_{j'=1}^{k} e^{a_j-\tia_{j'} }  \sum_{s=1}^{ \xi_{k-j'+1}^{j'}-\la_{k-j+2}^j    } e^{\ep_1 s} \non \\ 
&-& \! e^{-m}\sum_{k=1}^{N-1}\sum_{j=1}^{k}\sum_{j'=1}^{k} e^{a_j-\tia_{j'}} \sum_{s=1}^{ \xi_{k-j'+1}^{j'}-\la_{k-j+1}^j    } e^{\ep_1 s} \,,
\eea
which agrees with proposition 5.22 in \cite{Negut:2008} (after some changes in notation). An important thing to note is that the $y_N$-independent terms only depend on $\ep_1$ and {\it not} on $\ep_2$, which is similar to the setting in \cite{Nekrasov:2009} (see also \cite{Mironov:2009x,Nekrasov:2010}).  It was shown in \cite{Negut:2008} that the instanton partition function for the $\cN=2^*$ theory with $k_N=0$ is (up to a prefactor) an eigenfunction of the quantum {\it trigonometric} Calogero-Sutherland model. Connections between eigenfunctions of quantum integrable systems and instanton partition functions in the presence of surface operators have also been studied in \cite{Awata:2009,Alday:2010,Maruyoshi:2010}. In particular, in \cite{Alday:2010} (see also \cite{Teschner:2010}) it was argued that the instanton partition function  for the $\cN=2^*$ theory  in the critical limit $\ep_2 \rar 0$ is an eigenfunction of the quantum  {\it elliptic} Calogero-Moser model. This result is more directly related to the setup in  \cite{Nekrasov:2009}.

Whereas instanton partition functions built from the character (\ref{surfchar}) are intimately connected with the affine $\sll(N)$ algebra the results in \cite{Negut:2008} are based on the ordinary $\sll(N)$ algebra. We will see in later sections that this fact has a natural explanation since the part of the affine conformal blocks independent of the worldsheet coordinate $z$ is constructed from descendants that only involve the zero-modes of the affine current, which span the ordinary $\sll(N)$ Lie algebra.

It is also possible to consider quivers with more that one $\SU(N)$ factor. Here we consider one of the simplest such models, the superconformal $\SU(N){\times}\SU(N)$ model with one matter multiplet of mass $m$ transforming in the bifundamental representation, $N$ multiplets with masses $\mu_i$ transforming in the fundamental representation of the first $\SU(N)$ factor and  $N$  multiplets with masses $\tilde{\mu}_i$ transforming in the fundamental representation of the second $\SU(N)$ factor.

The simplest class of terms are the ones with $k_i=n$ and $\tilde{k}_j=p$ (which arise when  only $\la_1^i = n$ and $\xi_1^j=p$ are non-zero). For terms of this type we find that the contribution to the instanton partition function is given by
\bea 
&& \sum_{n=0}^{\infty}\sum_{p=0}^{\infty} \frac{      (\frac{\tilde{\mu}_i}{\ep_1}{-}\frac{a_{i}}{\ep_1})_n \, (\frac{\tilde{a}_i}{\ep_1}{-}\frac{a_{i}}{\ep_1}{+}\frac{m}{\ep_1})_n  \, (\frac{\mu_{j+1}}{\ep_1}{-}\frac{\tilde{a}_{j}}{\ep_1}{+}\frac{\ep_2}{\ep_1}\lfloor \frac{j}{N}\rfloor {+}1)_p\,  (\frac{a_{j+1}}{\ep_1}{-}\frac{\tilde{a}_{j}}{\ep_1}{+}\frac{\ep_2}{\ep_1}\lfloor\frac{j}{N}\rfloor {+}1{-}\frac{m}{\ep_1})_p  }
{   (\frac{a_{i+1}}{\ep_1}-\frac{a_{i}}{\ep_1}+\frac{\ep_2}{\ep_1}\lfloor \frac{i}{N}\rfloor +1)_n \, n! \, (\frac{\tilde{a}_{j+1}}{\ep_1}-\frac{\tilde{a}_{j}}{\ep_1}+\frac{\ep_2}{\ep_1}\lfloor\frac{j}{N}\rfloor+1)_p \, p!} \non \\
&& \qquad \times \left[ \frac{(\frac{\tilde{a}_i}{\ep_1}-\frac{a_{i}}{\ep_1} - p+\frac{m}{\ep_1})_n   }{ (\frac{\tilde{a}_i}{\ep_1}-\frac{a_{i}}{\ep_1}+\frac{m}{\ep_1})_n   } \right]^{\de_{ij} } 
\left[ \frac{ (\frac{\tilde{a}_{j}}{\ep_1}-\frac{a_{j+1}}{\ep_1}-\frac{\ep_2}{\ep_1}\lfloor\frac{j}{N}\rfloor +\frac{m}{\ep_1})_n  }{  (\frac{\tilde{a}_{j}}{\ep_1}-\frac{a_{j+1}}{\ep_1}-\frac{\ep_2}{\ep_1}\lfloor\frac{j}{N}\rfloor -p+\frac{m}{\ep_1} )_n  } \right]^{\de_{i,j+1}} \!\!\! y_i^n \, \tilde{y}_j^p\,.
\eea
It is convenient to change notation for the masses 
\be
\frac{\tilde{\mu}_i}{\ep_1} \rar  \frac{\mu_{i+1}}{\ep_1} + \frac{\ep_2}{\ep_1}\lfloor {\frac{i}{N}}\rfloor +1\,,  \qquad  \qquad
\frac{\mu_{i+1}}{\ep_1} \rar  \frac{\tilde{\mu}_i}{\ep_1} - \frac{\ep_2}{\ep_1}\lfloor \frac{i}{N}\rfloor -1\,.
\ee
Using this notation the above expression becomes
\bea \label{quiverinst}
&& \sum_{n=0}^{\infty}\sum_{p=0}^{\infty} \frac{      (  \frac{\mu_{i+1}}{\ep_1}{-}\frac{a_{i}}{\ep_1}{+} \frac{\ep_2}{\ep_1}\lfloor \frac{i}{N}\rfloor {+}1)_n \, (\frac{\tilde{a}_i}{\ep_1}{-}\frac{a_{i}}{\ep_1}{+}\frac{m}{\ep_1})_n \,  (\frac{a_{j+1}}{\ep_1}{-}\frac{\tilde{a}_{j}}{\ep_1}{+}\frac{\ep_2}{\ep_1}\lfloor\frac{j}{N}\rfloor {+}1{-}\frac{m}{\ep_1})_p   \, (\frac{\tilde{\mu}_j}{\ep_1}{-}\frac{\tilde{a}_{j}}{\ep_1})_p }
{   (\frac{a_{i+1}}{\ep_1}-\frac{a_{i}}{\ep_1}+\frac{\ep_2}{\ep_1}\lfloor \frac{i}{N}\rfloor +1)_n \, n! \, (\frac{\tilde{a}_{j+1}}{\ep_1}-\frac{\tilde{a}_{j}}{\ep_1}+\frac{\ep_2}{\ep_1}\lfloor\frac{j}{N}\rfloor+1)_p \, p!} \non \\
&& \qquad \times \left[ \frac{(\frac{\tilde{a}_i}{\ep_1}-\frac{a_{i}}{\ep_1} - p+\frac{m}{\ep_1})_n   }{ (\frac{\tilde{a}_i}{\ep_1}-\frac{a_{i}}{\ep_1}+\frac{m}{\ep_1})_n   } \right]^{\de_{ij} } 
\left[ \frac{ (\frac{\tilde{a}_{j}}{\ep_1}-\frac{a_{j+1}}{\ep_1}-\frac{\ep_2}{\ep_1}\lfloor\frac{j}{N}\rfloor +\frac{m}{\ep_1})_n  }{  (\frac{\tilde{a}_{j}}{\ep_1}-\frac{a_{j+1}}{\ep_1}-\frac{\ep_2}{\ep_1}\lfloor\frac{j}{N}\rfloor -p+\frac{m}{\ep_1} )_n  } \right]^{\de_{i,j+1}} \!\!\! y_i^n \, \tilde{y}_j^p\,.
\eea
In this form it is easy to see that the terms with $p=0$ or $n=0$ reduce to (\ref{Nf4Z0}) with $(a_i,\mu_i,\tilde{\mu}_i) = (a_i,\mu_i,\tilde{a}_i+m)$  and $(a_i,\mu_i,\tilde{\mu}_i) = (\tilde{a}_i,a_i-m,\tilde{\mu}_i)$, respectively.

\setcounter{equation}{0} 
\section{Affine $\sll(2)$ and surface operators in $\SU(2)$ gauge theories} \label{ssl2}

In  \cite{Alday:2010} it was argued that the  instanton  partition  function in an $\SU(2)$ quiver gauge theory with  a full surface operator insertion  is  equal  to a   modified  version of an affine $\sll(2)$  conformal block. In this  section we  review  and  check   this  proposal, showing  how  the analytical  results of  the previous  section  can  be reproduced  from  affine  conformal  blocks. We   consider the four- and five-point conformal blocks on the sphere and  the one-point conformal block  on  the  torus. These   are  associated to the  $\SU(2)$ theory with four flavours, the $\SU(2){\times}\SU(2)$ quiver with a bifundamental hypermultiplet and two flavours in each $\SU(2)$ factor,  and  the  $\cN=2^*$ $\SU(2)$  gauge theory which has one adjoint hypermultiplet. In  order to  fix  our  conventions, we  start by reviewing some  basic  facts  about  the affine $\sll(2)$ Lie algebra.

The  commutation relations that define the untwisted affine $\sll(2)$ Lie algebra (usually denoted $\widehat{\sll}(2)$ or $A_1^{(1)}$) are given  by
\be
[J_n^0,J_m^0] = \frac{k}{2} \, n\, \de_{n+m,0} \,,\quad 
[J_n^0,J_m^{\pm}] = \pm J_{n+m}^{\pm}  \,, \quad
[J_n^+,J_m^-] = 2J_{n+m}^0 + k\,  n\, \de_{n+m,0} \,.
\ee
Primary states with respect to this algebra satisfy $J_0^0| j \rb=j | j \rb$ and are  annihilated by 
\be
J^-_{1+n} |j\rb =J^0_{1+n}|j\rb=J^+_{n}|j\rb=0 \qquad  \,  (n=0,1,2,\ldots)\,,
\ee
which implies that
\be
\, \lb j | J^+_{-1+n} =\lb j | J^0_{-1+n}=\lb j | J^-_{n}=0 \qquad \; (n=0,-1,-2,\ldots)\,.
\ee
We denote  the corresponding  primary field $V_j(x,z)$, where $x$ is an isospin variable and $z$ is the worldsheet coordinate. The action of  the  generators  on a  primary  field  can  be expressed  in  terms  of  differential  operators:
\be \label{JV}
[ J_n^A,V_j(x,z) ] = z^n D^A V_j(x,z) \,,
\ee
where
\be
D^+ = 2\, j\, x  - x^2 \pa_x  \,, \qquad 
D^0 = -x\pa_x + j \,, \qquad 
D^- = \pa_x \,,
\ee
which satisfy\footnote{Since $[J_n^A,[J^B_m,V_j]] =z^{n+m}D^B D^A V_j$, consistency 
of (\ref{JV}) 
implies that  $[ [J_n^A,J^B_m],V_j(x,z) ] = -z^{n+m} [D^A,D^B] V_j(x,z)$.}
\be
[D^0,D^{\pm}] = \mp D^{\pm}  \,, \qquad
[D^+,D^-] = -2D^0\, .
\ee
The descendants of a primary  state, $\lb j|$, are  denoted $\lb {\bf n, A};j |$, where  
\be
\lb {\bf n, A};j | = \lb j | J_{n_1}^{A_1} \cdots J_{n_\ell}^{A_\ell}\,,
\ee
and we define the level $ n=\sum_i n_i$ and charge $ \Ups= \sum_i A_i$.  For  later  reference, we  recall  that for  the affine  $\sll(2)$ algebra the matrix of inner products of descendants (usually called the  Gram or Shapovalov matrix) satisfies 
\be 
X_{ {\bf n, A} ; {\bf n}', \bf{A}'}(j)  = \lb {\bf n, A};j |  {\bf n', A'};j \rb\propto\delta_{ n, n'}\delta_{ \Ups,\Ups'}\,,
\ee
i.e.~it is a block-diagonal matrix where each block contains only  descendants with  given values for the level $n$  and  charge $\Ups$.

\subsection{Four-point conformal block on the sphere}

Our first example is the four-point conformal block on the sphere.    Following the  proposal in \cite{Alday:2010}, this should  equal, up  to a prefactor,  the instanton partition  function  for  the $\SU(2)$ theory with $N_f=4$  with  a full surface  operator insertion. In our conventions,
\be\label{4ptat}
Z_{\rm inst}=(1-z)^{2j_2(-j_3+k/2)}\lb j_1| V_{j_2}(1,1)  \cK(x,z) V_{j_3}(x,z) |j_4 \rb \,,
\ee
where $\cK(x,z)$ is  an  operator  defined  as
\be \label{KAT}
\cK(x,z)=\exp\left[-\sum_{n=1}^{\infty}\frac{1}{2n-1} \left(z^{n-1} x \, J^-_{1-n} + \frac{z^n}{ x} J^+_{-n}\right)\right] .
\ee
The  insertion of the  $\cK(x,z)$  operator is  not  strictly  necessary  for  the case  of  the  four-point block on the sphere. It is possible  to  reproduce the instanton  partition function  also without $\cK$, by considering a small modification of the  dictionary below. However, since the $\cK$ operator  is  crucial  when matching the higher-point  conformal blocks  to instanton partition functions in quiver   gauge theories,  we  will insert a $\cK$ operator, following the prescription in \cite{Alday:2010} (note that the expression for $\cK$ written in \cite{Alday:2010} is equal to $\cK(1,1)$ in our notation).

In  order  to  reproduce  the  results  of the  previous  section,  we  consider  the following  standard  decomposition of  the  conformal  block\footnote{Here and  in all similar expressions in  the  following, we  omit  the  three-point  factors. (In (\ref{4ptb}) the $
 \lb j_1| V_{j_2}(1,1)  | j \rb \lb j | V_{j_3}(x,z) |j_4 \rb$ factors in  the  denominator on the right hand side are implicit.)}
\bea\label{4ptb}
 \!\!\! \!\!\! &&\lb j_1| V_{j_2}(1,1)  \cK(x,z) V_{j_3}(x,z) |j_4 \rb \non \\
 \!\!\! \!\!\! &=& \!\!\! \!\!\!   \sum_{{\bf n, A}; {\bf n}', {\bf A}'}
 \lb j_1| V_{j_2}(1,1)  |{\bf n}, {\bf A}; j \rb  X^{-1}_{ {\bf n, A} ; {\bf
n}', \bf{A}'}(j)  \lb {\bf n}', {\bf A}';j | \cK(x,z) V_{j_3}(x,z) |j_4 \rb \,.
\eea
Before we proceed with the computation of this object we would like to point out that it is also possible to  reproduce the instanton  partition functions (also for the quiver cases) by using a slightly different insertion, namely 
\bea\label{4ptbKdag}
 \!\!\! \!\!\! &&\lb j_1| V_{j_2}(1,1)  \cK^{\dag}(1,1) V_{j_3}(x,z) |j_4 \rb \non \\
 \!\!\! \!\!\! &=& \!\!\! \!\!\!   \sum_{{\bf n, A}; {\bf n}', {\bf A}'}
 \lb j_1| V_{j_2}(1,1)  \cK^{\dag}(1,1)  |{\bf n}, {\bf A}; j \rb  X^{-1}_{ {\bf n, A} ; {\bf
n}', \bf{A}'}(j)  \lb {\bf n}', {\bf A}';j |  V_{j_3}(x,z) |j_4 \rb \,.
\eea
where
\be \label{Kdag}
\cK^{\dag}(x,z)=\exp\left[\sum_{n=1}^{\infty}\frac{1}{2n-1} \left(\frac{z^{-n+1}}{ x} \, J^+_{n-1} + z^{-n} x J^-_{n}\right)\right] .
\ee
This operator will be important in section \ref{sslN}, but here we continue to use the expressions ({\ref{4ptb}) and (\ref{KAT}).

Note that affine and conformal invariance imply that 
\be \label{3ptN2}
\lb j_1| V_{j_2}(x,z)   | j_3 \rb \propto x^{j_2 + j_3 - j_1} z^{\Delta_1 - \Delta_2 - \Delta_3} \,,
\ee
where $\De_i$ denotes the conformal dimension of the $i$th state. Using this result and (\ref{JV}), it is possible to  compute  the conformal block (\ref{4ptb}) perturbatively (cf.~e.g.~\cite{Awata:1992}). The result  is a series with only positive powers of $z$ but both positive and negative powers of $x$. However,  the  power  of $x$ in the denominator  can  only  be equal to  or  smaller than the  power of  $z$  in the numerator. Two  limiting cases are  thus  given  by the $z$-independent terms and the subset  of terms containing only  powers of  $\frac{z}{x}$. We start by considering the $z$-independent terms. These arise  from descendants in the internal channel of  the form  $(J_0^-)^n | j \rb$.  Note that for  descendants of this type, the  Gram matrix  is    diagonal   and  can be trivially  inverted.  These  terms  thus  lead to the following contribution   
\bea\label{cbx}\non
&&\!\! \sum_{n=0}^{\infty}\frac{  \lb j_1| V_{j_2}(1,1)   (J_0^-)^n | j \rb \lb j | (J_0^+)^n  e^{ -x J_0^-} V_{j_3}(x,z) |j_4 \rb }{\lb j | (J_0^+)^n   (J_0^-)^n | j \rb }\\
&=&\!\! \sum_{n=0}^{\infty}\frac{(-x)^n}{n!} \frac{  (j_1-j_2-j)_n  (- j - j_4+j_3 )_{n} }{ (-2j)_{n}  } \,,
\eea
where we used that $\lb j | (J_0^+)^n  \cK(x,z)=\lb j | (J_0^+)^n  e^{ -x J_0^-}$ (see appendix \ref{appmix} for some additional details).
In a similar way, the  terms that involve  only powers  of  $\frac{z}{x}$ can be  computed.   These  arise from descendants in the  internal channel of  the form  $(J_{-1}^+)^n | j \rb$ that  have  a diagonal Gram matrix and lead to the contribution
\bea\label{cbxz}\non
&&\!\! \sum_{n=0}^{\infty}\frac{  \lb j_1| V_{j_2}(1,1)   (J_{-1}^+)^n | j \rb \lb j | (J_1^-)^n e^{ -\frac{z}{x} J_{-1}^+} V_{j_3}(x,z) |j_4 \rb }{\lb j | (J_1^-)^n   (J_{-1}^+)^n | j \rb }\\
 &=& \!\! \sum_{n=0}^{\infty} \left(-\frac{z}{x}\right)^n\frac{1}{n!} \frac{  (j-j_1-j_2)_n  ( j + j_4+j_3 -k)_{n} }{ (2j-k)_{n}  } \,,
\eea
 where we   used that $\lb j |(J_1^-)^n  \cK(x,z)=\lb j |(J_1^-)^n e^{ -\frac{z}{x}J_{-1}^+}$. 
 The expressions (\ref{cbx}) and  (\ref{cbxz}) are both hypergeometric functions of the form ${}_2F_1(A ,B ; C ;y)$. 
 
Next we describe the dictionary between the variables on the two sides of the conjectured equality (\ref{4ptat}). The worldsheet coordinate $z$ and the isospin  coordinate $x$   are  related to the   instanton expansion parameters $y_1$ and $y_2$  as  
\be \label{xyz}
y_1=x\,, \qquad \qquad y_2 = \frac{z}{x} \,.  
\ee
Note that this identification is consistent with the fact that the instanton partition function contains only positive powers of $y_1$, $y_2$.
The  momenta  of the  external  states of the  conformal  block are  related to  the  hypermultiplet masses, the  momentum  of  the  internal  state is  related to the   Coulomb modulus and  the level  of  the affine  algebra  is  related  to the deformation parameters. The  precise  dictionary is   
\bea\label{k4ptmap}\non
j_1&\!\!=\!\!&-\frac{\epsilon_1+\epsilon_2+\mu_1-\mu_2}{2\epsilon_1}\,,\quad \; \,  j_2=-\frac{2\epsilon_1+\epsilon_2+\mu_1+\mu_2}{2\epsilon_1}\,, \; \, \, j =-\frac{1}{2}+\frac{a_1}{\epsilon_1}\,, \\
 j_3&\!\!=\!\!&-\frac{2\epsilon_1+\epsilon_2-\tilde{\mu}_1-\tilde{\mu}_2}{2\epsilon_1}\,, \quad 
 j_4=-\frac{\epsilon_1+\epsilon_2+\tilde{\mu}_1-\tilde{\mu}_2}{2\epsilon_1}\,,\quad  k=-2-\frac{\epsilon_2}{\epsilon_1} \,.
\eea
 Using  the  dictionary  (\ref{xyz}), (\ref{k4ptmap}) one easily checks that (\ref{cbx}), (\ref{cbxz})  are equal to\footnote{Note that for  these terms  the  prefactor in (\ref{4ptat})   does  not give  any  contribution.}   the corresponding components of the instanton partition function,  (\ref{Nf4Z0}).

We  have also  analysed  the  terms  of the  conformal  block  of the form $x^n z$. Such terms arise from internal  states of the form  
\be
|1\rb=  J_{-1}^{+} (J_0^{-} )^{n+1}  | j \rb \,,   \qquad |2\rb=J_{-1}^{0} (J_0^{-} )^n  | j \rb \,,  \qquad |3\rb=J_{-1}^{-} (J_0^{-} )^{n-1}  | j \rb \,.
\ee
For  any $n\geq 1$ the  above  states  generate  a $3{\times}3$  sub-block  of the  Gram matrix\footnote{When $n=0$,  the block  reduces  to a $2{\times}2$  block.} and  the  $x^n z$ term  of  the conformal  block is given by
\be\label{4pmixk}
\sum_{r,s=1}^3
 \lb j_1| V_{j_2}(1,1)  |r \rb  X_{r,s}^{-1}  \lb s |\cK(x,z)  V_{j_3}(x,z) |j_4 \rb \,,
\ee   
where $X_{r,s}= \lb r | s\rb$ with $r,s=1,2,3$ is the relevant  block of the Gram  matrix. The expression (\ref{4pmixk}) can be  computed  by noting  that for the states considered it  can be shown that 
\be
\lb r |\cK(x,z)  =\lb r |\re^{-J^-_0 x -
z\left(\frac{1}{3}J_{-1}^{-}x+\frac{J_{-1}^{+}}{x}\right)}=\lb r |\re^{-J^-_0 x } \left[ 1+z   \left(  -\frac{J_{-1}^{+}}{x}+J^0_{-1}      \right)    \right],
\ee
where we made  use  of the Zassenhaus formula (\ref{Zassy}). Using  the  dictionary  (\ref{k4ptmap}) it can be shown that the infinite set of terms obtained from (\ref{4pmixk}) correctly reproduce the component  $Z_{\rm inst}^{(1)1,2}$ of the $\SU(2)$ instanton partition function, (\ref{nf41}).  Some details of the computation can be found in appendix \ref{appmix}.

\subsection{Five-point conformal  block on the sphere}

Our second example is the   five-point conformal block on the sphere. In this case we consider
\be
\lb j_1| \cV_{j_2}(1,1)  \cV_{j_3}(x,z) \cV_{j_4}(\tilde{x},\tilde{z})  |j_5 \rb \,,
\ee
where we introduced the notation
\be \label{KV}
\cV_j(x,z) =   \cK(x,z) V_{j}(x,z) \,.
\ee
In order  to  match the conformal  block  to  the results  for  the $\SU(2){\times}\SU(2)$  quiver gauge theory, we use  the  standard  decomposition
\be \label{kcb5}
 \sum_{{\bf p},{\bf p}',{\bf n},{\bf n}'}
\!  \lb j_1| \cV_{j_2}(1,1)   |{\bf p}; j \rb X^{-1}_{ \bf p; \bf
p'}(j)  \lb {\bf p}';j | \cV_{j_3}(x,z)  |{\bf n}; \tij \rb X^{-1}_{ \bf n; \bf
n'}(\tij)  \lb {\bf n}';\tij | \cV_{j_4}(\tilde{x},\tilde{z}) |j_5 \rb,
\ee
where  for brevity we  omitted the ${\bf A }$-type internal indices.  Note that $\cV_{j_2}$ can be replaced by $V_{j_2} $ since $\lb j_1 | \cK =\lb j_1 | $. As mentioned above it is also possible to use $ \cV_j(x,z) =   V_{j}(x,z) \cK^\dag(x,z)$, but here we continue to use (\ref{KV}). 
Let us  first focus  on  the  terms in (\ref{kcb5}) with ${\bf n=n'=0}$.  The non-trivial part is exactly  the  same four-point block  that we  considered  in the previous  section.   Summing the terms with $|{\bf p} ; j \rb=(J_0^-)^p | j \rb $ produces    
\be \label{4ptx5k}
\sum_{p=0}^{\infty} \frac{ (j_1-j_2-j)_p  (-j-\tij+j_3)_p}{ (-2j)_p } \frac{(-x)^p}{p!}\,.
\ee
Similarly,  summing the terms with $|{\bf p} ; j \rb=(J_{-1}^+)^p | j \rb$ gives 
\be\label{4ptxz5k}
\sum_{p=0}^{\infty} \frac{ (j-j_1-j_2)_p (j+\tij+j_3-k)_p}{p! (2j-k)_p  } \left(- \frac{z}{x} \right)^p.
\ee
Next we consider the  terms in (\ref{kcb5}) with ${\bf p=p'=0}$.  
The two families  of  internal states   $|{\bf n} ; \tij\rb=(J_0^-)^n | \tij \rb $ and $|{\bf n} ; \tij \rb=(J_{-1}^+)^n | \tij \rb$ give    
\be \label{4pb}
\sum_{n=0}^{\infty} \frac{ (j-j_3-\tij)_n  (j_4-j_5-\tij)_n}{n! (-2\tij)_n } \left(-\frac{\tilde{x}}{x}\right)^n,
\ee
and
\be\label{4ptc}
\sum_{n=0}^{\infty} \frac{ (\tij-j-j_3)_n (\tij+j_4+j_5-k)_n}{n! (2\tij-k)_n  } \left( -\frac{x\tilde{z}}{z \tilde{x} } \right)^n .
\ee
The above expressions (\ref{4ptx5k})-(\ref{4ptc}) are all hypergeometric functions.  
  These  four  hypergeometric  functions can  be  matched  to the  instanton computation  for the $\SU(2)\times\SU(2)$  quiver gauge theory (\ref{quiverinst}). Note that in (\ref{quiverinst}),  the  terms with $n=0$  sum to  two  hypergeometric  functions in $\tilde{y}_1$ and $\tilde{y}_2$, respectively, and the $p=0$ terms sum to two  hypergeometric  functions in $y_1$ and $y_2$, respectively.   The   map between the expansion parameters is  given by
\be
  -x= \tilde{y}_1 \,, \qquad - \frac{z}{x} = \tilde{y}_2\,, \qquad -\frac{\tilde{x}}{x}=y_1\,, \qquad -\frac{x\tilde{z}}{z\tilde{x}} = y_2\,,
\ee
 and the remaining dictionary is
 \bea\label{qmapk}\non
j_1&\!\! = \!\!&\frac{-\epsilon_1+\tilde{\mu}_1-\tilde{\mu}_2}{2\epsilon_1}\,,\qquad \,
j_2=-\frac{\tilde{\mu}_1+\tilde{\mu}_2}{2\epsilon_1}\,,\qquad \qquad \, 
j_3=-\frac{m}{\ep_1} \,,\\ 
j_4&\!\!=\!\!&\frac{\mu_1+\mu_2}{2\epsilon_1}\,, \qquad \qquad \quad  
j_5=\frac{-\epsilon_1+\mu_1-\mu_2}{2\epsilon_1} \\ 
j& \!\!=\!\!&-\frac{1}{2}+\frac{\tilde{a}_1}{\epsilon_1}\,, \qquad\qquad \quad
\tij =-\frac{1}{2}+\frac{a_1}{\epsilon_1} \,, \qquad\qquad \quad
k=-2-\frac{\epsilon_2}{\epsilon_1} \,.\non
\eea
From the expression (\ref{kcb5}) it is also possible to  correctly  reproduce  the terms in (\ref{quiverinst}) with both $p\neq 0$ and $n\neq 0$.  Using  the  notation
\bea\non
| 1,p; j \rb&=&(J_0^-)^p | j \rb \,, \qquad \, X_{(1,p)}(j)=\lb j|(J_0^+)^p(J_0^-)^p | j \rb \,, \\
 | 2,p; j \rb&=&(J_{-1}^+)^p | j \rb \,, \qquad X_{(2,p)}(j)=\lb j |(J_{1}^-)^p(J_{-1}^+)^p | j \rb \,,
\eea
we can  summarize  the  result by noting  that  that the generic  $n$, $p$ term in (\ref{quiverinst}),  is  equal  to the term in the  conformal  block of the form 
\be 
 \lb j_1| \cV_{j_2}(1,1)   | r,p; j \rb X^{-1}_{(r,p)}(j)  \lb r,p; j| \cV_{j_3}(x,z)  |s,n; \tij\rb X^{-1}_{ (s,n)}(\tij)  \lb s,n;\tij | \cV_{j_4}(\tilde{x},\tilde{z})  |j_5 \rb ,
\ee
where $r,s=1,2$. In conclusion, we should stress that the operator $\cK$ was crucial for the match to the instanton result. (It is also possible to reproduce the instanton result by inserting $\cK$ only next to  $V_{j_3}$, rather than next to  both $V_{j_3}$ and $V_{j_4}$ as we did above.)

\subsection{One-point conformal block on the torus}

Our final example is the one-point conformal block on the torus:
\be\label{torus}
 Z_{\rm 1pt}^{\cK}=\sum_{{\bf n; A},{\bf n}';\bf{A}'} z^n x^\Ups 
 \lb {\bf n, A};j |\cK(x,z) V_{j_1}(x,z)   |{\bf n}', {\bf A}'; j \rb X^{-1}_{ \bf n, A ; {\bf
n}', \bf{A}'}(j) \,,
\ee
where  $\lb {\bf n, A};j | = \lb j | J_{n_1}^{A_1} \cdots J_{n_\ell}^{A_\ell}$,  $n=\sum_i n_i$ and $\Ups= \sum_i A_i$. It was argued  in \cite{Alday:2010} that, for this case, the only effect of the insertion of the $\cK$ operator is a prefactor:
\be\label{toruspre}
Z_{\rm 1pt}^{\cK}=(1-x-\frac{z}{x})^{-j_1}Z_{\rm 1pt} \,,
\ee
where $Z_{\rm 1pt}$ is  the one-point conformal block on the  torus  without  the $\cK$ operator, i.e. 
\be\label{torusnok}
 Z_{\rm 1pt}=\sum_{{\bf n; A},{\bf n}';\bf{A}'} z^n x^\Ups 
 \lb {\bf n, A};j |V_{j_1}(x,z)   |{\bf n}', {\bf A}'; j \rb X^{-1}_{ \bf n, A ; {\bf
n}', \bf{A}'}(j) \,.
\ee
We  have  checked  that the relation (\ref{toruspre}) is  satisfied for the terms that  arise from internal states of the form $(J_0^-)^n | j \rb$ or $(J_{-1}^+)^n | j \rb$. In the  following we  therefore  focus  on the conformal block  without the $\cK$ operator insertion, (\ref{torusnok}).   As above, we  first  compute the $z$-independent terms that arise from the  internal states  $(J_0^-)^n | j \rb$. These lead to the result
\bea\label{torusx}
 \sum_{n=0}^{\infty}  x^n  \sum_{\ell=0}^n   \Big(\!\!\ba{c} n \\ \ell \ea \!\! \Big)  \frac{(-1)^\ell}{\ell!} \frac{(-j_1)_\ell(j_1+1)_\ell}{(-2j)_\ell } \,.
\eea
The terms involving powers of $\frac{z}{x}$ are due  to the $(J_{-1}^+)^n | j \rb$ internal  states and gives
\bea \label{torusxz}
 \sum_{n=0}^{\infty}  \left(\frac{z}{x}\right)^n  \sum_{\ell=0}^n   \Big(\!\!\ba{c} n \\ \ell \ea \!\! \Big)  \frac{(-1)^\ell}{\ell!} \frac{(-j_1)_\ell(j_1+1)_\ell}{(2j-k)_\ell } \,.
\eea
From  the general result 
\bea\label{hprel}
(1-x)^{A-1}{}_2 F_1(A,C-B;C;x)=\sum_{n=0}^{\infty}x^n\sum_{\ell=0}^n\Big(\!\!\ba{c} n \\ \ell \ea \!\! \Big) \frac{(-1)^\ell}{\ell !}\frac{A_\ell B_\ell}{ C_\ell} \,,
\eea
one sees that (\ref{torusx}) and (\ref{torusxz}) can be written as 
\bea
(1-x)^{j_1}{}_2 F_1(1+j_1,-2j+j_1;-2j;x)\,,
\eea
and
\bea
\left(1-\frac{z}{x}\right)^{j_1}{}_2 F_1\left(1+j_1,2j-k+j_1;2j-k;\frac{z}{x}\right).
\eea
It  then follows  that by using (\ref{toruspre}) and considering  the  dictionary 
\be\label{wimap2}
x=y_1\,, \qquad\frac{z}{x}=y_2\,, \qquad j_1=-\frac{m}{\epsilon_1}\, ,\qquad j =-\frac{1}{2}+\frac{a}{\epsilon_1}\,,\qquad k=-2-\frac{\epsilon_2}{\epsilon_1} \,,
\ee  
the conformal  block precisely reproduce the result for the $\cN=2^*$ $\SU(2)$ gauge theory (\ref{N2*Z0}) that we obtained in the previous section.

\subsection{Liouville theory and surface  operators in $\SU(2)$ gauge theories}\label{m2m5}

In this section we explore the relation between surface operators  that in the $6d$ language arise from $2d$ and $4d$ defects.  For  $\SU(2)$ gauge theories there is only one Levi-type of surface  operator and even though the $2d$ and $4d$ defects are different objects one can investigate if the instanton partition function is sensitive to the difference.

Consider  the $\SU(2)$ gauge  theory  with  four  flavours in the presence of a (simple) surface operator arising from a $2d$ defect.   According  to the  proposal in \cite{Alday:2009b}, the  instanton  partition  function  should equal   the  Liouville  conformal  block  with four non-degenerate primaries  and  one degenerate field.  We  have   verified  that, up to a prefactor, the Liouville conformal  block
\bea
\lb \alpha_1| V_{\alpha_2}(1)  V_{-\frac{b}{2}}(x)V_{\alpha_3}(z) |\alpha_4 \rb \,,
 \eea
is indeed  in  agreement  with  the  instanton computation for  the $\SU(2)$ $N_f=4$ gauge  theory in the presence of a surface operator arising from a $4d$ defect
that we  described  in  section \ref{sinst}.\footnote{The  details  of  this  computation  are collected in Appendix \ref{liouville}.}    (For the pure $\SU(2)$ theory a similar check was performed in \cite{Maruyoshi:2010}.)

As   mentioned above, for the $\SU(2)$ $N_f=4$ gauge theory, the instanton partition function can also be  reproduced  from an $\widehat{\sll}(2)$ conformal  block {\it without} the $\cK$ operator insertion. This implies that the result we  have just  described is  in agreement with the  Zamolodchikov-Fateev result \cite{Zamolodchikov:1986},  that  shows that the Liouville five-point conformal  block  with a degenerate  field insertion is  equal (up to a prefactor)  to  the standard four-point $\widehat{\sll}(2)$ conformal  block. 

For the conformal $\SU(2){\times}\SU(2)$ quiver gauge theory we considered above one expects a relation between the five-point  $\widehat{\sll}(2)$ conformal  block (with the $\cK$ operator insertion) and the Liouville conformal block
\bea
\lb \alpha_1| V_{\alpha_2}(1)  V_{-\frac{b}{2}}(x)V_{\alpha_3}(z) V_{-\frac{b}{2}}(\chi)V_{\alpha_4}(\zeta)  |\alpha_5 \rb \,.
 \eea
Using the standard decomposition, one obtains to lowest order the same structure that we found above involving  four hypergeometric functions, but in this case it  is not straightforward to find a relation between the two expressions (possibly one can find a map if one allows for mixing between  internal/external momenta and masses/Coulomb moduli).

This indicates that already at the quiver level the instanton partition function is sensitive to the difference between $2d$ and $4d$ defects. This is perhaps not surprising since in the M-theory 
setup the  surface operator arising from a $4d$  defect involves an M5-brane, whereas the surface operator arising from a $2d$ defect involves an M2 brane \cite{Alday:2010}\footnote{This is reminiscent of the situation for the $\cN=4$ $\SU(N)$ theories where surface operators can be constructed both using D3-branes \cite{Gukov:2006} and also using D7-branes \cite{Buchbinder:2007}.  }. Also note that  already for the four-point block the map is not of the form one naively would have expected  since (as can be shown) $\al_{1,2}$ are not mapped to $j_{1,2}$. The map between full and simple surface operators deserves to be further studied.


\setcounter{equation}{0}
\section{Affine $\sll(N)$ and surface operators in $\SU(N)$ gauge theories}  \label{sslN}

In this section we make a proposal for how to extend the $\widehat{\sll}(2)$ analysis discussed in section \ref{ssl2}  to $\widehat{\sll}(N)$. Compared to the extension of the $\SU(2)$/Liouville results in \cite{Alday:2009a} to the $\SU(N)$/Toda results in \cite{Wyllard:2009}, one difference is that we will be able to do computations for arbitrary $N$, since the affine $\sll(N)$ algebra is simpler than the $\cW_{N}$ algebra.

We start by recalling some facts about the $\sll(N)$ Lie algebra. The $\gl(N)$ Lie algebra can be defined in terms of $N{\times} N$ matrices $E^{IJ}$ whose only non-zero entry is a 1 at position $(I,J)$. These matrices  satisfy the commutation relations
\be \label{glN}
[E^{IJ},E^{KL}] = \de^{JK} E^{IL} - \de^{LI}E^{KJ}\,.
\ee
For later purposes it will be convenient to use a composite index $I=(0,i)$ where $i=1,\ldots,N-1$.   The generators of the $\sll(N)$ subalgebra of $\gl(N)$ are given by e.g.
\be
E^i \equiv H^i= (E^{ii}-E^{i-1,i-1})/2\,, \quad \;  E^{i+}\equiv E^{i0}\,, \quad  \; E^{i-}\equiv E^{0i}\,, \quad \; E^{i l} \; (i\neq l) \,.
\ee
 The commutation relations of these generators can be obtained from (\ref{glN}).

In a completely analogous convention, the generators, $J_n^a$, of the affine $\sll(N)$ Lie algebra (usually denoted $\widehat{\sll}(N)$ or $A_{N-1}^{(1)}$) are (here $i,l=1,\ldots,N-1$ and $n\in \ZZ$)
\be
J_n^i\,,  \qquad \quad  J^{i+}_n\,,  \qquad \quad J^{i-}_n\,,  \qquad \quad J^{i l}_n \quad( i \neq l)\,.
\ee
Most of the commutation relations are the obvious ones induced from those of $\sll(N)$. The non-trivial ones involving the level $k$ are\footnote{This follows from the general result $[J_n^A,J_m^B]= f^{ABC}J^C_{n+m} + k\, m \, \de_{m+n,0}\, \ka^{ab}$, where $\ka^{ab}$ is the Killing form of $\sll(N)$, which in terms of the $E^{IJ}$ $N{\times}N$ matrices can be written $\ka^{IJ;KL} = \tr(E^{IJ}E^{KL})$.}: 
\bea
 [J_n^{i},J_m^{j}] &\!=\!& \frac{k}{4}\,n\,A_{ij}\, \de_{n+m,0} \,, \qquad   [J_n^{i+},J_m^{i-}] \,= \, k\,n\,\de_{n+m,0} + 2 \sum_{s=1}^{i} J_{n+m}^s  \,, \non \\ 
{} [J_n^{i l}, J_m^{li}]&\!=\!& k\,n\,\de_{n+m,0} + \sum_{s=l+1}^{i} J_{n+m}^s \,,
 \eea
where $A_{ij}=\lb e_i,e_j\rb$ is the Cartan matrix. For a given value of the level $k$, primary states, $|j\rb$, with respect to the $\widehat{\sll}(N)$ algebra are labelled by a vector $j$ in the $N-1$ dimensional root/weight space of $\sll(N)$. Such a vector, $j$, can be expanded as 
\be \label{jLa}
j = \sum_{i=1}^{N-1} j^i \La_i \,,
\ee
 where $\La_i$ are the fundamental weights of $\sll(N)$ (see appendix \ref{ALie} for a summary of our Lie algebra conventions). A primary state satisfies 
\be
J_0^i | j \rb= j^i | j \rb\,, \quad \; J_0^{i+} | j \rb =0 \,, \quad \; J_0^{i l} | j \rb =0 \;\; (i>l)\,, \quad \; J_n^{A} | j \rb =0 \;\; (n>0)\,.
\ee
The  primary field, $V_j(x,z)$, corresponding to the primary state $|j\rb$ satisfies
\be \label{JVN}
[ J_n^a,V_j(x,z) ] = z^n D^a V_j(x,z)\,,
\ee
where $z$ denotes the worldsheet coordinate and $x$ denotes a collection of isospin variables. 
The relation (\ref{JVN})  generalises the $\widehat{\sll}(2)$ result (\ref{JV}). In general, the $D^a$'s in (\ref{JVN})  depend on $N(N{-}1)/2$ isospin variables (which equals the number of positive roots of $\sll(N)$); see e.g.~\cite{Rasmussen:1996} for some examples. However, as will become clear later, in the cases of main interest to us the primary fields appearing in the commutators with the $J_n^a$'s always satisfy $j = \chi =  \ka \La_{N-1}$ (or $j = \chi =  \ka \La_{1}$).  
For such special primary fields we will now argue that a smaller number of isospin variables is sufficient.

There is a known realisation of $\gl(N)$ in terms of differential operators acting on the space spanned by $\{x_i\}$ where $i=1,\ldots,N{-}1$ (this is the space of smallest dimension where $\gl(N)$ can act)\footnote{This realisation is perhaps better known in terms of $N{-}1$ oscillators and dates back to \cite{Okubo:1974}.}. More precisely, in this realisation the generators are
\be \label{glND}
D^{00}  = -2 \ka +\sum_i x_i \pa_{x_i} \,, \quad D^{i+} \equiv D^{i0}= -x_i D^{00} \,, \quad D^{i-}\equiv D^{0i} = \pa_{x_i}  \,, \quad
D^{il} = -x_i \pa_{x_l}\,.
\ee
These generators satisfy the commutation relations 
\be \label{glNDcomm}
[D^{IJ},D^{KL}] = -\de^{JK} D^{IL} + \de^{LI}D^{KJ} \,,
\ee
i.e.~the same commutation relations as the $N{\times} N$ matrices $E^{IJ}$ in (\ref{glN}), but with the opposite sign on the right hand side. Note that the generators in (\ref{glND}) depend on {\it one} parameter, $\ka$. 

For the restriction to $\sll(N)$ we use the same notation as before, i.e. 
\be
D^i = (D^{ii}-D^{i-1,i-1})/2\,, \quad \; D^{i+}\equiv D^{i0}\,, \quad \;  D^{i-}\equiv D^{0i}\,, \quad \; D^{il} \; \; (i\neq l) \,.
\ee
In terms of the $D$'s, a highest weight representation of $\sll(N)$ is obtained from 
\be
D^i v_{ j}(x) = j^i \, v_j(x)\,, \qquad D^{i+} v_j(x)  =0 \,, \qquad D^{il} v_j(x)  =0 \quad (i>l)\,.
\ee
In particular, when $2\ka$ takes the integer value $n$ we find a finite-dimensional representation space (module) spanned by $x_1^{n_1} \cdots x_{N-1}^{n_{N-1}}$ with $0 \le \sum n_i \le n = 2\ka$. The highest weight is easily found from the above conditions:  the second condition implies that $v_j(x)=x_1^{n_1} \cdots x_{N-1}^{n_{N-1}}$ with $\sum n_i = n $, and the third condition then implies that $v_j(x)= x_{N-1}^{n}$. Finally, we have $D^i   x_{N-1}^n = -\frac{n}{2} \, \de_{i,N-1} \, x_{N-1}^n$, i.e.~the representation corresponds to the highest weight\footnote{This is  analogous to the situation in the $\cW_{N}$ algebra, where a semi-degenerate state with momentum $\al=\ka \La_{N-1}$ becomes degenerate when $\al=- n b \La_{N-1}$.} $-\frac{n}{2} \La_{N-1}= -\ka \La_{N-1}$. (Similarly, the lowest weight is $v_j(x)= 1$, satisfying $D^i   v_j = \frac{n}{2} \, \de_{i,1} \, v_j$, corresponding to $\frac{n}{2} \La_{1}= \ka \La_{1}$.)

Taken together, these facts indicate that the $\sll(N)$ generators extracted from the $D$'s in (\ref{glND}) can be used as $D^a$'s in (\ref{JVN}) when the primary field $V_j$ has a $j$ of the form $\ka \La_{N-1}$ ($\ka \La_{1}$). This proposal is very natural from the point of view of the conjecture in \cite{Alday:2010} since a full surface operator in an $\SU(N)$ gauge theory depends on precisely $N-1$ variables.

\subsection{Four-point conformal block on the sphere}

Next we turn to explicit examples and checks of the proposed relation between (slightly modified) affine $\sll(N)$ conformal blocks and instanton partition functions in  $\SU(N)$ quiver gauge theories with a full surface operator insertion. Our first example is the four-point conformal block on the sphere: 
\be \label{cbN}
 \sum_{{\bf n, A};{\bf n}', {\bf A}'}
 \lb j_1| \cV_{\chi_2}(1,1)  |{\bf n}, {\bf A}; j \rb X^{-1}_{ \bf n,  A; 
n',  A' }(j)  \lb {\bf n}', {\bf A}';j |  \cV_{\chi_3}(x,z) |j_4 \rb \,,
\ee
where $j, j_1$, and $j_4$ now denote arbitrary $N-1$ dimensional vectors and $\chi_i = \ka_i \La_{N-1}$ (or $\chi_i = \ka_i \La_{1}$). As before,  $X_{ \bf n, A; \bf n', A'}(j)=    \lb {\bf n, A}; j |{\bf n', A'} ; j \rb $ and  $|{\bf n',A'} ; j \rb$ is a descendant of the primary state  $|j\rb$:
\be
|{\bf n},{\bf A};j \rb =  J_{-n_1}^{A_1} \cdots J_{-n_\ell}^{A_\ell} | j \rb \,.
\ee
We conjecture that the expression (\ref{cbN})  is equal (up to a prefactor) to the instanton partition function of the $\cN=2$  $\SU(N)$ theory with $N_f=2N$ and a full surface operator insertion. Based on the results in section \ref{ssl2} we expect that 
\be
\cV_{\chi_i}(x,z) = V_{\chi_i}(x,z)  \cK^\dagger (x,z) \qquad \mathrm{or} \qquad \cV_{\chi_i}(x,z) = \cK (x,z)  V_{\chi_i}(x,z)  \,.
\ee
We propose the following two natural generalisations of (\ref{KAT}) and (\ref{Kdag}):
\bea \label{KN}
\cK(x,z) &=& \exp\bigg[- \sum_{n=1}^{\infty} \frac{1}{2n-1} \bigg(\sum_{i=1}^{N-1} \frac{z^{n-1}x^i}{i} J^{i-}_{1-n} - \sum_{i<l} \frac{z^{n-1}}{l-i} \frac{x_l}{x_i} J_{1-n}^{il} \non \\ && \qquad \;  \qquad  \qquad  \qquad  + \sum_{i=1}^{N-1} \frac{1}{i}\frac{z^n}{x_i}J^{i+}_{-n} - \sum_{i<l}   \frac{z^{n}}{l-i}\frac{x_i}{x_l}  J_{-n}^{li} \bigg)  \bigg],
\eea
and 
\bea \label{KdagN}
\cK^\dag(x,z) &\!=\!& \! \exp\bigg[ \sum_{n=1}^{\infty} \frac{1}{2n-1} \bigg(\sum_{i=1}^{N-1} \frac{1}{i} \frac{z^{n-1}}{x^i} J^{i+}_{n-1} - \sum_{i<l} \frac{z^{-n+1}}{l-i} \frac{x_i}{x_l} J_{1-n}^{li} \non \\ && \qquad \quad  \qquad  \qquad + \sum_{i=1}^{N-1}  \frac{1}{i} z^{-n} x_i J^{i-}_{n} - \sum_{i<l}     \frac{z^{-n}}{l-i}\frac{x_l}{x_i}  J_{-n}^{il} \bigg) \bigg] .
\eea
These expression are a guess based on the conjecture for $\widehat{\sll}(2)$ \cite{Alday:2010} together with  $\sll(N)$ covariance. Note that the expressions in the exponent for different values of $n$ commute. Also note that the zero mode part, i.e. the piece involving only $J_0^A$, can also be written in  factorised form
\be
\cK_0 = e^{-J_0^{1-}} e^{J_0^{12 }} \cdots  e^{J_0^{N-2,N-1 }} \,, \qquad \qquad \cK_0^\dag = e^{J_0^{1+ }}  e^{-J_0^{21 }} \cdots  e^{-J_0^{N-1,N-2 }} \,.
\ee
We should stress that in our explicit examples below we will only check parts of these expressions. It would be desirable to have further checks and a better understanding of $\cK^\dag$ and $\cK$.

Note that affine and conformal invariance implies that 
\be \label{3ptN}
\lb j_1| V_{\chi_2}(x,z)   | j_3 \rb \propto z^{\Delta_1 - \Delta_2 - \Delta_3} \prod_{i=1}^{N-1}  x_i^{2(\frac{\ka_2}{N} - \lb h_{i+1},j_3\rb +\lb h_{i+1},j_1\rb)   } .
\ee
Here $h_i$ ($i=1,\ldots,N$) are the weights of the fundamental representation (see appendix \ref{ALie} for a summary of our Lie algebra conventions). This result can be derived by inserting $J_0^i$ into the three-point function and using
\bea
j_1^i \lb j_1| V_{\chi_2}(x,z)   | j_3 \rb&=& \lb j_1| J_0^i V_{\chi_2}(x,z)   | j_3 \rb =  \lb j_1| ([J_0^i, V_{\chi_2}(x,z)] +   V_{\chi_2}(x,z)J_0^i ) | j_3 \rb \non \\
&=& (\half [D^{ii}-D^{i-1,i-1}] + j_3^i)\lb j_1| V_{\chi_2}(x,z)   | j_3 \rb\,.
\eea 
(The result of this argument shows that the exponents of the $x_i$ can be written in terms of $j_r^i$ where $r=1,2,3$, i.e~in terms of the components of $j_{1,2,3}$ in the expansion (\ref{jLa}); using   the conventions  in appendix \ref{ALie} the exponents can then be written in the above form.)  
Note that (\ref{3ptN}) reduces to (\ref{3ptN2}) when $N=2$.

Using (\ref{JVN}) and (\ref{3ptN}) the four-point conformal block can be computed perturbatively. The result is a series with only positive powers of $z$ but both positive and negative powers of the $x_i$. 

As a first example we consider the $z$-independent terms. Among all possible such terms, there are $N{-}1$  subsets, each involving a power series in one particular combination of the $x_i$'s, that only receive contributions from one type of descendants. More precisely, these subsets involve only terms of the form $x_1^n$ or only $(\frac{x_{i}}{x_{i-1}})^n$ for a fixed $i$ with $i=2,\ldots,N-1$, and  arise from descendants $\lb {\bf n}',{\bf A}';j |$ involving only $J_0^{1+}$ or only  $J_0^{i,i-1}$ for a fixed $i$  with $i=2,\ldots,N-1$.

The only non-zero $X_{ \bf n, A; \bf n', A'}(j)$ involving the relevant descendants are 
\be
 \lb j| (J_0^{1+})^n  (J_0^{1-})^n | j \rb = n! (-2j^1)_n(-1)^n \,,  \quad \lb j| (J_0^{i,i-1})^n  (J_0^{i-1,i})^n | j \rb =  n! (-2j^{i})_n(-1)^n,
\ee
where for the second expression $i=2\ldots,N-1$; note that the two results fit together nicely in one  $\sll(N)$ covariant expression.

Let us first focus on the $x_1^n$ terms and compute the four-point block {\it without} the insertion of $\cK$ or $\cK^\dag$, i.e.~we use $\cV_{\chi_i} = V_{\chi_i} $. In this case we find 
\bea\label{try1}
&& \frac{  \lb j_1| V_{\chi_2}(1,1)   (J_0^{1-})^n | j \rb \lb j | (J_0^{1+})^n V_{\chi_3}(x,z) |j_4 \rb }{\lb j | (J_0^{1+})^n   (J_0^{1-})^n | j \rb }  \non \\ 
&=&  \frac{(-2\frac{\ka_2}{N}-2\lb h_2,j_1\rb +2\lb h_2,j\rb)_n(-2\frac{\ka_3}{N} +2\lb h_1,j_4\rb-2\lb h_1,j\rb)_n }{n! (-2j^1)_n}  (x_1)^n \,.
\eea
Summing the $x_1^n$ terms then leads to the  hypergeometric function ${}_2F_1(A,B;C;x)$ with 
\be \label{NABC}
A = -2\frac{\ka_2}{N} \,{+}2\lb h_2,j\rb{-}2\lb h_2,j_1\rb \,,\! \quad B=-2\frac{\ka_3}{N}\, {+}2\lb h_1,j_4\rb{-}2\lb h_1,j\rb\,, \! \quad C= -2j^1\,.
\ee
Similarly if we use $\cV_{\chi_i}(x,z) = \cK (x,z)  V_{\chi_i}(x,z)$ instead, we find
\bea \label{try2}
&& \frac{  \lb j_1| V_{\chi_2}(1,1)  (J_0^{1-})^n | j \rb \lb j   | (J_0^{1+})^n e^{-x_1 J_0^{1-}} V_{\chi_3}(x,z) |j_4 \rb }{\lb j | (J_0^{1+})^n   (J_0^{1-})^n | j  \rb }  \non \\ 
&=&  \frac{(-\frac{2\kappa_2}{N}-2\lb h_2 , j_1\rb + 2 \lb h_2, j \rb)_n(\frac{2\kappa_3}{N}-2\lb h_1 , j_4\rb + 2 \lb h_2, j \rb)_n }{n! (-2j^1)_n}  \left(- x_1 \right)^n .
\eea
Note that the full $\cK$ (\ref{KN}) is not needed here; only a part contributes. Finally if we use $\cV_{\chi_i}(x,z) = V_{\chi_i}(x,z)  \cK^\dagger (x,z)$ we find
\bea \label{try3}
&& \frac{  \lb j_1| V_{\chi_2}(1,1) e^{J_0^{1+}} (J_0^{1-})^n | j \rb \lb j   | (J_0^{1+})^n V_{\chi_3}(x,z) |j_4 \rb }{\lb j | (J_0^{1+})^n   (J_0^{1-})^n | j  \rb }  \non \\ 
&=&  \frac{(\frac{2\kappa_2}{N}+2\lb h_2 , j_1\rb - 2 \lb h_1, j \rb)_n(-\frac{2\kappa_3}{N}+2\lb h_1 , j_4\rb - 2 \lb h_1, j \rb)_n }{n! (-2j^1)_n}  \left(- x_1 \right)^n ,
\eea
where again the full $\cK^\dag$ is not needed. 
The above expressions should be compared to the $y_1^n$ terms in the instanton partition function (\ref{Nf4Z0}), which is also of hypergeometric form. By equating the denominators we see that (up to a constant) we should equate $j^1 \propto a_1-a_2$. This result in turn implies that  $\lb h_i , j\rb \propto a_i$ (for $i=1,2$ and again up to a constant). Since the $y_1^n$ terms in (\ref{Nf4Z0}) only involve $a_1$ in the numerator, it seems that only (\ref{try3}) can equal the instanton result.  For this reason we will use insertions of $\cK^\dag$ in the remainder of this section.

For the $\widehat{\sll}(2)$ four-point conformal block all three of the above possibilities could be matched to the instanton result (using minor modifications in the dictionary). It is easy to see why: for $\sll(2)$ we have that $h_1=-h_2$. For the $\SU(2)$ quiver theories, as pointed out in section \ref{ssl2}, insertions of either $\cK$ or $\cK^\dag$ are needed. It is conceivable that also in higher rank theories one can use $\cK$ insertions provided that one uses suitable conventions, but here we will use $\cK^\dag$ since it results in expressions that can be matched to the instanton results in a straightforward and natural way.

Before we proceed with the computation, let us mention another property of the insertions of $\cK$ and $\cK^\dag$ that may turn out to be important. As can be seen from the above expressions (\ref{try1}), (\ref{try2}) the effect of the insertion of $\cK$ is to turn ${}_2F_1(A,B;C;x_1)$ into  ${}_2F_1(A,C-B;C;-x_1)$. Similarly, as can be seen from  (\ref{try3})  the $\cK^\dag$ insertion results in ${}_2F_1(C-A,B;C;-x_1)$.  Hypergeometric functions satisfy various identities, such as
\be
{}_2F_1(A,C-B;C;x)= (1-x)^{-A} {}_2F_1(A,B;C;-\frac{x}{1-x})\,,
\ee
therefore a possible alternative to the insertion of $\cK^\dag$ ($\cK$) might be to change variables instead (or to pick a different solution of the hypergeometric differential equation), but we will not pursue this idea here. 

Returning to the computation we find that in addition to (\ref{try3}) we also have 
\bea \label{try3i}
\!\!\!\!&&\!\! \frac{  \lb j_1| V_{\chi_2}(1,1)  e^{-J_0^{i,i-1}}(J_0^{i-1,i})^n | j \rb \lb j | (J_0^{i,i-1})^n V_{\chi_3}(x,z) |j_4 \rb  }{\lb j | (J_0^{i,i-1})^n   (J_0^{i-1,i})^n | j \rb } \non \\
\!\!\!\!&=& \!\! \frac{(\frac{2\kappa_2}{N}+2\lb h_{i+1} , j_1\rb - 2 \lb h_{i}, j \rb)_n (-\frac{2\kappa_3}{N}+2\lb h_{i} , j_4\rb - 2 \lb h_{i}, j \rb)_n}{n! (-2j^{i})_n} \left( -\frac{x_{i}}{x_{i-1}}\right)^n \!\!\!,
\eea
where $i=2,\ldots,N-1$. Let us emphasize again that in the above expressions the full $\cK^\dag$ (\ref{Kdag}) is not needed. It is easy to see that all $J_{n}^A$ dependence drops out for $n>0$. To see that also most of the $J_0^A$ dependence drops out requires a bit more thought. 

For the $z$-dependent part, terms of the form $(\frac{z}{x_{N-1}})^n$ also only receive contributions from one source, namely from the the descendants involving only $J_{1}^{N-1,-}$.
The relevant component of the Gram matrix $X_{ \bf n, A; \bf n', A'}(j)$  is
\be
 \lb j| (J_1^{N-1,-})^n  (J_{-1}^{N-1,+})^n | j \rb = n! (2\,[\sum_{i=1}^{N-1} j^i]-k)_n(-1)^n \,.
 \ee
 Using this result we find:
\bea \label{zxterms}
\!\!\!\!\!\!\!\!&&\!\! \frac{   \lb j_1| V_{\chi_2}(1,1)   e^{J_{1}^{N-1,-}} (J_{-1}^{N-1,+})^n | j \rb \lb j | (J_1^{N-1,-})^nV_{\chi_3}(x,z) |j_4 \rb  }{\lb j | (J_1^{N-1,-})^n   (J_{-1}^{N-1+})^n | j \rb }  \non \\
\!\!\!\!\!\!\!\!&=&\!\!  \frac{(\frac{2\kappa_2}{N}+2\lb h_{1} , j_1\rb {-} 2 \lb h_{N}, j \rb{-}k)_n (-\frac{2\kappa_3}{N}+2\lb h_{N} , j_4\rb {-} 2 \lb h_{N}, j \rb)_n}{n! ( 2\,[\sum_{i=1}^{N-1} j^i]-k   )_n}  \left( -\frac{z}{x_{N-1}}\right)^n \!\!\! .
\eea
Note that $j^i = \lb e_i,j\rb = \lb u_i-u_{i+1},j\rb $. This implies that $\sum_{i=1}^{N-1} j^i = -\lb u_N-u_1,j\rb$ which shows that (\ref{zxterms}) fits together nicely with the other results (\ref{try3}), (\ref{try3i}) provided that one identifies $u_{i+N}=u_i$.

The above results should be compared with the instanton result (\ref{Nf4Z0}). We propose that
\be \label{yis}
y_1 = x_1 \,, \qquad y_{i+1}=\frac{x_{i+1}}{x_{i}} \quad (1\le i \le N{-}2) \,, \qquad y_{N}=\frac{z}{x_{N-1}} \,.
\ee
 Non-trivial evidence for this identification comes from the fact that, as is easy to see, all terms in the expansion of the conformal block can be written as a power series with only positive powers of the $y_i$'s (the instanton result (\ref{Nf4Z0}) is also a power series in $N$ variables $y_i$ with only positive powers). 
To  match  the  denominators of (\ref{try3}), (\ref{try3i}) to (\ref{Nf4Z0}) we identify
\be \label{jisa}
 j^i = -\frac{1}{2}+\frac{a_i-a_{i+1}}{2\ep_1}\,, \qquad \quad  \quad k=-N-\frac{\epsilon_2}{\epsilon_1} \,.
\ee
Given  that $\sum_{i=1}^{N}a_i=0$,  it  follows  from  the  above  formula that
\be
2\,[\sum_{i=1}^{N-1} j^i]-k= \frac{a_1-a_{N}}{\ep_1} +1+\frac{\epsilon_2}{\epsilon_1} \,,
\ee
and  thus  the  denominator in (\ref{zxterms}) also agrees  with the  instanton result.  Defining $a=\sum_i^{N}a_iu_i$, the relation (\ref{jisa}) can also be written in various other ways 
\be
  \lb e_i , j\rb = \frac{1}{2}\lb e_i,\frac{a}{\ep_1}-\rho\rb \,, \quad \;  \lb u_i , j\rb = \frac{1}{2}\lb u_i,\frac{a}{\ep_1}-\rho\rb  \,, \quad \; \lb h_i , j\rb = \frac{1}{2}\lb h_i,\frac{a}{\ep_1}-\rho\rb\,.
\ee
We also propose the following dictionary for the masses
\be \label{mmap}
 \frac{\tilde{\mu}_i}{2\ep_1} = -\frac{\ka_3}{N}+\lb h_i , j_4+\frac{\rho}{2}\rb \,, \qquad  \frac{\mu_{i}}{2\ep_1} = \frac{\ka_2}{N} + \lb h_{i} ,j_1+\frac{\rho}{2} \rb \,,   
\ee
which leads to complete agreement between (\ref{Nf4Z0}) and the results in this section. 
(Note that the $\widehat{\sll}(2)$ version of the map (\ref{mmap}) is slightly simpler than the one used in (\ref{k4ptmap}) which arose from an expression where $\cK$ was used instead of $\cK^\dag$.)

One can also compute corrections to the above expressions. One particular class of such corrections involve terms of the form $y_i^n\, y_l$ with $l\neq i$. The first thing to note is that when $l\neq i\pm 1$ there is only one possible way to obtain such terms. This result agrees with the structure of the instanton expansion (\ref{nf41}). When  $i,l$ belong to the range $2,\ldots,N-1$ the contributing descendants are of the form 
\be
\lb j | (J_0^{i,i-1})^n  (J_0^{l,l-1}) \,.
\ee
The corresponding terms in the conformal block  are easily computed. When $l\neq i\pm 1$, one finds (\ref{try3i}) multiplied by
\be
 \frac{(\frac{2\kappa_2}{N}+2\lb h_{l+1} , j_1\rb - 2 \lb h_{l}, j \rb)(-\frac{2\kappa_3}{N}+2\lb h_{l} , j_4\rb - 2 \lb h_{l}, j \rb)}{ (-2j^{l})} \left( -\frac{x_{l}}{x_{l-1}}\right) \!\!\!,
\ee
This result is easily seen to agree with the instanton result (\ref{nf41}) (when $l\neq i\pm 1$) using the maps (\ref{yis}), (\ref{jisa}), and (\ref{mmap}). When $l = i\pm 1$ the situation is slightly more involved. We have checked that the $x_1^n x_2$ terms correctly reproduce the instanton result (\ref{nf41}). This computation is sensitive to other terms in $\cK^\dag$ besides the ones appearing in the zeroth order analysis; some formul\ae{} are collected in appendix \ref{appmix}.


\subsection{Five-point conformal  block on the sphere}

Our next example is the  five-point conformal block (for brevity we suppressed the ${\bf A, A'}$-type labels): 
\be \label{cb5N}
 \sum_{{\bf n},{\bf n'},{\bf m},{\bf m}'}
 \lb j_1| \cV_{\chi_2}(1,1) |{\bf n}; j \rb X^{-1}_{ \bf n; \bf
n'}(j)  \lb {\bf n}';j |\cV_{\chi_3}(x,z)  |{\bf m};\tilde{\jmath}\rb X^{-1}_{ \bf m; \bf
m'}(\tilde{\jmath})  \lb {\bf m}';\tilde{\jmath}| \cV_{\chi_4}(\tilde{x},\tilde{z})   |j_5 \rb\,,
\ee
where $\cV_{\chi_i}(x,z) = V_{\chi_i}(x,z) \cK^\dagger(x,z)$ and  we  have  inserted  two  complete sets  of  states $|{\bf n} ; j \rb$ and $|{\bf m} ; \tilde{\jmath} \rb$. 
Using (\ref{JVN}) the conformal block can be computed perturbatively in  powers  of $x_i, z$ and  $\tilde{x}_i, \tilde{z}$. Just like in the $\widehat{\sll}(2)$ analysis, certain terms with ${\bf m=m'=0}$ or  ${\bf n=n'=0}$ can easily be computed.  The  terms with ${\bf m=m'=0}$ lead to hypergeometric functions of the type determined in the four-point analysis above:
\bea
\!\!\!\!\!\!\!\!&& \frac{  \lb j_1| V_{\chi_2}(1,1) e^{J_0^{1+}} (J_0^{1-})^n | j \rb \lb j   | (J_0^{1+})^n V_{\chi_3}(x,z) |\tij \rb }{\lb j | (J_0^{1+})^n   (J_0^{1-})^n | j  \rb }  \non \\ 
\!\!\!\!\!\!\!\!&=&  \frac{(\frac{2\kappa_2}{N}+2\lb h_2 , j_1\rb - 2 \lb h_1, j \rb)_n(-\frac{2\kappa_3}{N}+2\lb h_1 , \tij \rb - 2 \lb h_1, j \rb)_n }{n! (-2j^1)_n}  \left(- x_1 \right)^n , \non \\
\!\!\!\!\!\!\!\!\!\!&&\!\!\frac{  \lb j_1| V_{\chi_2}(1,1)  e^{-J_0^{i,i-1}}(J_0^{i-1,i})^n | j \rb \lb j | (J_0^{i,i-1})^n V_{\chi_3}(x,z) | \tij  \rb  }{\lb j | (J_0^{i,i-1})^n   (J_0^{i-1,i})^n | j \rb } \non \\
\!\!\!\!\!\!\!\!\!\!&=&\!\!  \frac{(\frac{2\kappa_2}{N}+2\lb h_{i+1} , j_1\rb {-} 2 \lb h_{i}, j \rb)_n (-\frac{2\kappa_3}{N}+2\lb h_{i} , \tij \rb {-} 2 \lb h_{i}, j \rb)_n}{n! (-2j^{i})_n} \left( -\frac{x_{i}}{x_{i-1}}\right)^n \!\!\!,
\eea
and
\bea
\!\!\!\!\!\!\!\!&&\!\!\frac{   \lb j_1| V_{\chi_2}(1,1)   e^{J_{1}^{N-1,-}} (J_{-1}^{N-1,+})^n | j \rb \lb j | (J_1^{N-1,-})^nV_{\chi_3}(x,z) | \tij  \rb  }{\lb j | (J_1^{N-1,-})^n   (J_{-1}^{N-1+})^n | j \rb }  \\
\!\!\!\!\!\!\!\!&=& \!\! \frac{(\frac{2\kappa_2}{N}{+}2\lb h_{1} , j_1\rb - 2 \lb h_{N}, j \rb-k)_n (-\frac{2\kappa_3}{N}+2\lb h_{N} , \tij \rb {-} 2 \lb h_{N}, j \rb)_n}{n! ( 2\,[\sum_{i=1}^{N-1} j^i]-k   )_n}  \left( -\frac{z}{x_{N-1}}\right)^n\!\!.
\eea
Similarly, when ${\bf n=n'=0}$  we  obtain  hypergeometric functions from
\bea
\!\!\!\!\!\!&& \!\!\frac{  \lb j | V_{\chi_3}(x,z) e^{\frac{1}{x_1}J_0^{1+}} (J_0^{1-})^n | \tij \rb \lb \tij   | (J_0^{1+})^n V_{\chi_4}(\tilde{x},\tilde{z}) |j_5 \rb }{\lb \tij| (J_0^{1+})^n   (J_0^{1-})^n |\tij \rb }  \non \\ 
\!\!\!\!\!\!&=&  \!\!\frac{(\frac{2\kappa_3}{N}+2\lb h_2 , j\rb - 2 \lb h_1, \tij\rb)_n(-\frac{2\kappa_4}{N}+2\lb h_1 , j_5\rb - 2 \lb h_1, \tij \rb)_n }{n! (-2\tij^1)_n}  \left(- \frac{\tilde{x}_1}{x_1} \right)^n , \non \\ 
\!\!\!\!\!\!&&\!\! \frac{  \lb j | V_{\chi_3}(x,z) e^{-\frac{x_{i-1}}{x_{i}}J_0^{i,i-1}}(J_0^{i-1,i})^n |\tij\rb \lb\tij | (J_0^{i,i-1})^n V_{\chi_4}(\tilde{x},\tilde{z}) |j_5 \rb  }{\lb j | (J_0^{i,i-1})^n   (J_0^{i-1,i})^n | j \rb }\non  \\
\!\!\!\!\!\!&=&\!\!  \frac{(\frac{2\kappa_3}{N}+2\lb h_{i+1} , j\rb - 2 \lb h_{i}, \tij \rb)_n (-\frac{2\kappa_4}{N}+2\lb h_{i} , j_5\rb - 2 \lb h_{i},\tij \rb)_n}{n! (-2\tij^{i})_n} \left( -\frac{\tilde{x}_{i} \, x_{i-1}}{\tilde{x}_{i-1}\,x_{i}}\right)^n \!\! ,
\eea
and
\bea
\!\!\!\!\!\!\!\!\!\!&&\!\!\frac{   \lb j | V_{\chi_3}(x,z)  e^{\frac{x_{N-1}}{z}J_{1}^{N-1,-}} (J_{-1}^{N-1,+})^n | \tij \rb \lb\tij| (J_1^{N-1,-})^nV_{\chi_4}(\tilde{x},\tilde{z}) |j_5 \rb  }{\lb j | (J_1^{N-1,-})^n   (J_{-1}^{N-1+})^n | j \rb }  \non \\
\!\!\!\!\!\!\!\!\!\!&=&  \!\!\frac{(\frac{2\kappa_3}{N}+2\lb h_{1} , j\rb {-} 2 \lb h_{N}, \tij \rb{-}k)_n (-\frac{2\kappa_4}{N}+2\lb h_{N} , j_5\rb {-} 2 \lb h_{N}, \tij \rb)_n}{n! ( 2\,[\sum_{i=1}^{N-1} \tij^i]-k   )_n}  \left( -\frac{\tilde{z} \, x_{N-1}}{z \,\tilde{x}_{N-1}}\right)^n \!\!.
\eea
The  precise dictionary which equates the above expressions to the 
instanton result (\ref{quiverinst}) is 
\bea 
y_1 &\!\!=\!\!& -x_1 \,, \qquad y_{i+1}=-\frac{x_{i+1}}{x_{i}} \quad (1\le i \le N{-}2) \,, \qquad y_{N}=-\frac{z}{x_{N-1}}\,, \non \\
\tilde{y}_1 &\!\!=\!\!& -\frac{\tilde{x}_1}{x_1} \,, \! \qquad \tilde{y}_{i+1}=-\frac{\tilde{x}_{i+1} \, x_{i}}{\tilde{x}_{i}\,x_{i+1}} \quad (1\le i \le N{-}2) \,, \quad \tilde{y}_{N}=-\frac{\tilde{z} \, x_{N-1}}{z \,\tilde{x}_{N-1}}\,,
\eea
and
\bea
\lb h_i , j \rb&\!\!=\!\!&\frac{1}{2}\lb h_i,\frac{a}{\ep_1}-\rho\rb\,, \quad \;\; \; \lb h_i , \tij\rb=\frac{1}{2}\lb h_i,\frac{\tilde{a}}{\ep_1}-\rho\rb \,, \qquad \qquad \; \,  \; k=-N-\frac{\epsilon_2}{\epsilon_1} \,, \non \\
\frac{{\mu}_i}{2\ep_1} &\!\!=\!\!&\frac{\ka_2}{N} + \lb h_{i} ,j_1
+\frac{\rho}{2} \rb  \,,  \quad \;
  \frac{\tilde{\mu}_{i}}{2\ep_1}=-\frac{\ka_4}{N} + \lb h_i ,j_5+\frac{\rho}{2}\rb \,, \quad \; \frac{m}{2\ep_1}=-\frac{\kappa_3}{N}\,.
\eea

It is also possible to compare terms of the form $y_i^n\,\tilde{y}_l^p$. The new ingredient is the cross-terms   
\be
 \lb {\bf n}';j | \cV_{\chi_3}(x,z)  |{\bf m};\tilde{\jmath}\rb\,.
\ee
To illustrate how the above computations are affected consider first the case when $i,l$ lie in the range $2,\ldots,N-2$. In this case the cross terms are
 \be
 \lb j | (J_0^{i,i-1})^n V_{\chi_3}(x,z) (J_0^{l-1,l})^p | \tilde{\jmath}\rb\,.
\ee
Now if $i=l$ then $J_0^{i+1,i}$ and $J_0^{l,l+1}$ do not commute which complicates the computation. Similarly, if $i=l+1$ then although  $J_0^{l+2,l+1}$ and  $J_0^{l,l+1}$ commute, they both act non-trivially on $x_l$ which again affects the calculation. However, apart from these two special cases, it is easy to see that the computation essentially factorises in the sense the coefficient in front of $y_i^n\,\tilde{y}_l^p$ is simply the product of the coefficient in front of $y_i^n$ in the expansion of the above hypergeometric function times the coefficient in front of $\tilde{y}_l^p$ in the expansion of the other hypergeometric function. This is precisely the structure we found in the instanton expression (\ref{quiverinst}).

\subsection{One-point conformal block on the torus}

Our final example is the one-point block on the torus:
\be 
  \sum_{{\bf n; A},{\bf n}';\bf{A}'} \!\! z^n (\prod_{i=1}^{N-1} x_i^{\Ups_i} )
 \lb {\bf n, A};j   | V_{\chi_1}(x,z) \cK^\dag(x,z)   |{\bf n}', {\bf A}'; j \rb X^{-1}_{ \bf n,A ; {\bf
n}',\bf{A}'}(j)  \,,
\ee
where  $\lb {\bf n, A};j | = \lb j | J_{n_1}^{A_1} \cdots J_{n_\ell}^{A_\ell}$ and $n=\sum_i n_i$. The coefficients $\Ups_i$ are determined as follows: a generator $J_n^{i l}$ in $\lb {\bf n, A};j | $ contributes $+1$ to $\Ups_i$ and $-1$ to $\Ups_l$, whereas $J_n^{i\pm}$ contributes $\pm 1$ to $\Ups_i$.  As for  the $\sll(2)$ case,  we assume that the only effect of the    $\cK^\dag$  operator is  the  introduction  of  a prefactor, and we therefore focus on the perturbative  expansion of the above conformal block without the $\cK^\dag$ insertion. As in previous examples, we start by computing  the $z$-independent terms. The $x_1^n$ terms arise from expressions of the form: 
\be 
 \frac{\lb j | (J_0^{1+})^n V_{\chi_1}(x,z)  (J_0^{1-})^n   |j \rb}{\lb j | (J_0^{1+})^n (J_0^{1-})^n   |j \rb}    =\sum_{\ell=0}^n   \Big(\!\!\ba{c} n \\ \ell \ea \!\! \Big)  \frac{(-1)^\ell}{\ell!} \frac{(-2\frac{\kappa_1}{N})_\ell(2\frac{\kappa_1}{N}+1)_\ell}{(-2j^1)_\ell } \,.
\ee
Similarly, the $(x_i/x_{i-1})^2$ terms  (for $i=2,\ldots,N-1$) arise from 
\be 
 \frac{\lb j | (J_0^{i,i-1})^n V_{\chi_1}(x,z)  (J_0^{i-1,i})^n   |j \rb}{\lb j | (J_0^{i,i-1})^n(J_0^{i-1,i})^n   |j \rb}    = \sum_{\ell=0}^n   \Big(\!\!\ba{c} n \\ \ell \ea \!\! \Big)  \frac{(-1)^\ell}{\ell!} \frac{(-2\frac{\kappa_1}{N})_\ell(2\frac{\kappa_1}{N}+1)_\ell}{(-2j^{i})_\ell } \,.
\ee
One can also compute the terms involving $( z/ x_{N-1})^n$:
\be \label{1ptxzN}
\frac{\lb j | (J_1^{N-1,-})^n V_{\chi_1}(x,z)  (J_{-1}^{N-1,+})^n   |j \rb}{\lb j | (J_1^{N-1,-})^n  (J_{-1}^{N-1,+})^n   |j \rb}   = \sum_{\ell=0}^n   \Big(\!\!\ba{c} n \\ \ell \ea \!\! \Big)  \frac{(-1)^\ell}{\ell!} \frac{(-2\frac{\kappa_1}{N})_\ell(2\frac{\kappa_1}{N}+1)_\ell}{(2[\sum_{i}^{N-1}j^i]-k)_\ell }\,.
\ee
Using the  formula (\ref{hprel}) it  follows that the  terms  discussed above contribute  as 
\bea
(1-x_1)^{2\frac{\kappa_1}{N}}{}_2 F_1(1+2\frac{\kappa_1}{N},-2j^1+2\frac{\kappa_1}{N};-2j^1;x_1)\,,
\eea
\bea
(1-\frac{x_{i+1}}{x_{i}})^{2\frac{\kappa_1}{N}}{}_2 F_1(1+2\frac{\kappa_1}{N},-2j^{i+1}+2\frac{\kappa_1}{N};-2j^{i+1};\frac{x_{i+1}}{x_{i}})\,,
\eea
and
\bea
(1-\frac{z}{x_{N-1}})^{2\frac{\kappa_1}{N}}{}_2 F_1(1+2\frac{\kappa_1}{N},2[\sum_{i}^{N-1}j^i]-k+2\frac{\kappa_1}{N};2[\sum_{i}^{N-1}j^i]-k;\frac{z}{x_{N-1}}) \,.
\eea
By using the identifications  (\ref{yis}) together with the dictionary 
\bea\label{wimap}\non
\kappa_1=-N\frac{m}{2\ep_1} ,\quad  j^i = -\frac{1}{2}+\frac{a_i-a_{i+1}}{2\ep_1} ,\quad k=-N-\frac{\epsilon_2}{\epsilon_1}
 \eea
we find that (up to a prefactor) these expressions are  equivalent to  the  instanton   partition  function in the $\cN=2^*$ $\SU(N)$ gauge theory where the corresponding terms take the form (\ref{N2*Z0}).

%
\setcounter{equation}{0}
\section{Asymptotically free $\SU(N)$ gauge theories and affine $\sll(N)$}  \label{snas}

So far we have only discussed conformal $\cN=2$ quiver gauge theories. But as we discuss in this section it is also possible to treat non-conformal (asymptotically free) $\cN=2$ theories. 

The extension of  the $\SU(2)$ AGT relation to non-conformal theories was carried out in \cite{Gaiotto:2009b}. In this paper Gaiotto conjectured that the instanton partition function for the pure $\SU(2)$ theory can be obtained via
\be \label{Gai}
Z_{\rm inst} = \lb \De,\La|\De,\La\rb \,,
\ee
where the state $|\De,\La\rb$ should satisfy 
\be \label{Gconds}
L_1 |\De,\La\rb = \La  |\De,\La\rb \,,\qquad L_n |\De,\La\rb = 0\quad(n\ge2)\,.
\ee
In an important further development \cite{Marshakov:2009} it was shown that the Gaiotto state $|\De,\La\rb$ is a particular state in the Verma module (thereby proving its existence):
\be \label{MM}
|\De,\La\rb = \sum_{Y} \La^{n} Q_{\De}^{-1}(1^n; Y) | Y,\De\rb \,,
\ee
where $Y$ denotes a partition (Young tableau) $\ell^{n_\ell}   \cdots 2^{n_2}1^{n_1}  $, where $n=|Y|$ is the number of boxes in $Y$,  $| Y ,\De\rb   $ denotes the descendant $(L_{-\ell})^{n_\ell} \cdots (L_{-2})^{n_2} (L_{-1})^{n_1}|\De\rb$ of the primary state $|\De\rb$ with conformal dimension $\De$, and $Q_{\De}(Y,Y')=\lb Y ,\De | Y' ,\De\rb   $ is the inner product of descendants (usually called the Gram or Shapovalov matrix) with inverse $Q^{-1}_{\De}$.  
When combining (\ref{MM}) with (\ref{Gai}) it follows that 
\be \label{Qinv}
Z_{\rm inst} =  \sum_{n=0}^{\infty} Q^{-1}_{\De}( 1^n; 1^n) \La^{2n} \,.
\ee
(Note that it follows from (\ref{Gconds}) that the only $\lb \De,Y|$ that have non-zero inner product with  $|\De,\La\rb$ are $\lb \De, 1^n|$). 

The  result (\ref{Qinv}) can also be obtained from the AGT relation by sending the masses to infinity in a conformal $\SU(2)$ theory \cite{Marshakov:2009,Alba:2009}, and was proven in \cite{Hadasz:2010}. The extension to higher rank $\SU(N)$ theories was discussed in \cite{Taki:2009}. 

The addition of simple surface operators to non-conformal theories is also possible: as in the conformal cases one inserts degenerate states in the  dual $2d$ CFT.
In \cite{Awata:2009} it was shown that for the $\SU(2)$ theory with a
(simple) surface operator, the dual conformal block satisfies a
differential equation (the same differential equation was also found
earlier in \cite{Braverman:2004a}, which reflects the fact that for the
non-conformal $\SU(2)$ gauge theory   surface operators obtained by $2d$ and $4d$ defects 
seems to be  associated to the same  instanton partition function).
Further aspects were studied in the recent paper \cite{Maruyoshi:2010}.

Our goal is to generalise the above construction to the non-conformal $\SU(N)$ theories with a full surface operator insertion. In other words, we want to find analogues of (\ref{Gai})-(\ref{Qinv}) in the module of the affine $\sll(N)$ algebra. We should point out that the construction below is in agreement with a result proven in the first paper in \cite{Braverman:2004a}. In particular,  (\ref{GJconds}) and (\ref{GJcondsN}) correspond to what is called a Whittaker vector in \cite{Braverman:2004a}. However, here we use the language of \cite{Gaiotto:2009b,Marshakov:2009} which is more familiar to physicists. We first study the rank one case.

\subsection{Pure $\SU(2)$} \label{puresu2}
As in previous sections,  we label the descendants of the primary state $|j\rb$ by
\be
|{\bf n},{\bf A};j \rb =  J_{-n_1}^{A_1} \cdots J_{-n_\ell}^{A_\ell} | j \rb \,,
\ee
where we put $J_{-n}^A$ to the left of $J_{-n'}^{A'}$ if $n>n'$ or if $A<A'$ and $n=n'$. 
We also define the matrix (denoted $X_{ {\bf n, A} ; {\bf n}', \bf{A}'}(j)$ in previous sections)
\be
Q_j({\bf n},{\bf A};{\bf n'},{\bf A'}) =  \lb {\bf n},{\bf A};j |{\bf n}',{\bf A}';j \rb \,.
\ee
The following set of descendants will play an important role in what follows
\be
|n,p;j \rb = (J_{-1}^{+})^p  (J_{0}^{-})^n |j\rb \,.
\ee
We denote the corresponding diagonal component of the inverse of the matrix $Q_j$, i.e.~$Q^{-1}_j$, by $Q^{-1}_j(n,p;n,p)$. 

We propose that the instanton expansion of the pure $\SU(2)$ theory in the presence of a (full) surface operator can be obtained from 
\be \label{prop}
 Z_{\rm inst}=\sum_{n=0}^{\infty} \sum_{p=0}^{\infty} Q_j^{-1}(n,p;n,p) \, x^n \left(\frac{z}{x}\right)^p.
\ee
This expression is the analogue of (\ref{Qinv}) when a full surface operator is present. 
To test this proposal, we first consider the terms containing only $x$. Since, $|n,0;j \rb $  only has a non-zero inner product with its conjugate:
\be 
Q_j(n,0;n,0)= \lb j | (J_0^+)^n   (J_0^-)^n | j \rb  = n! (-2j)_n (-1)^n  \,,
\ee
we find that 
\be
Q_j^{-1}(n,0;n,0) = [  Q_j(n,0;n,0)  ]^{-1} = \frac{1}{ n! (-2j)_n (-1)^n  }\,,
\ee
which inserted into our proposal (\ref{prop}) leads to
\be \label{Z0x}
\sum_{n=0}^{\infty} \frac{(-1)^n }{  n! (-2j)_n } x^n \,.
\ee
Similarly, one can also consider the terms involving only powers of $\frac{z}{x}$. In this case, since  $|0,p;j \rb $  only has non-zero inner product with its conjugate: 
\be
Q_j(0,p;0,p)=\lb j | (J_1^-)^p   (J_{-1}^+)^p | j \rb   = p! (2j-k)_p (-1)^p  \,,
\ee
our proposal leads to
\be \label{Z0zx}
\sum_{n=0}^{\infty} \frac{(-1)^p }{  p! (2j-k)_p } \left( \frac{z}{x}\right)^p \,.
\ee
The above two expressions (\ref{Z0x}) and (\ref{Z0zx}) are in perfect agreement with the instanton result (\ref{Z0pure}) provided we identify
\be \label{nonconfdict}
j =  \frac{a}{\ep_1}-\frac{1}{2} \,, \qquad  k= -2 -\frac{\ep_2}{\ep_1}\,, \qquad x= -\frac{y_1}{(\ep_1)^2} \,, \qquad \frac{z}{x}= -\frac{y_2}{(\ep_1)^2} \,.
\ee

As a further check we consider all terms of the form $x^n \frac{z}{x}$. In this case there are three states that form a closed subset under the inner product involving $|1,n;j\rb$, namely  
\be
|\tilde{1}\rb=  J_{-1}^{+} (J_0^{-} )^{n}  | j \rb ,   \qquad |\tilde{2}\rb=J_{-1}^{0} (J_0^{-} )^{n-1}  | j \rb ,\qquad |\tilde{3}\rb=J_{-1}^{-} (J_0^{-} )^{n-2}  | j \rb .
\ee
The corresponding $3{\times}3$ block of $Q$ is 
\be \label{Q33}
\left( \ba{ccc} [k{-}2j{+}2n] M(n) & M(n)& 0 \\M(n) &\frac{k}{2} M(n{-}1) & -M(n{-1}) \\ 0 & -M(n{-1}) & [k{+}2j{-}2(n{-}2)] M(n{-}2) \ea \right)
\ee
where 
\be
M(n)\equiv {\lb j |  (J_0^{-} )^{n}     (J_0^{-} )^{n}   | j \rb }=(-2j)_n n! (-1)^n.
\ee
From the inverse of (\ref{Q33}) we find that 
\be
Q_j^{-1}(n,1;n,1) = -(-1)^n \frac{ (4 + 4 j + 4 k + 2 j k + k^2 - 6 n - 4 j n - 2 k n + 2 n^2) }{(2 j - 
   k) (2 + k) (2 + 2 j + k) n! (-2j)_{n} } \,,
\ee
which leads to
\be
-\frac{z}{x} \sum_{n=0}^{\infty} \frac{(-1)^n (4 + 4 j + 4 k + 2 j k + k^2 - 6 n - 4 j n - 2 k n + 2 n^2)}{(2 j - 
   k) (2 + k) (2 + 2 j + k)   n! (-2j)_n } x^n\,.
\ee
This result is again in perfect agreement with the instanton result (\ref{Z1pure}) provided that we use the identifications (\ref{nonconfdict}).

As in the case without surface operators it is also possible to construct the analogue of the  Gaiotto state (\ref{Gconds}). We denote the corresponding state  $|x,z;j\rb$ and demand that it should satisfy
\be \label{GJconds}
J_0^{+} |x,z;j\rb = \sqrt{x} \,|x,z;j\rb \,, \qquad J_1^{-} |x,z;j\rb = \sqrt{\frac{z}{x}}\, |x,z;j\rb \,, \qquad 
\ee
where all other $J_n^A$'s that annihilate $|j\rb$ also annihilate $|x,z;j\rb$. Finally, the analogue of (\ref{MM}) is
\be 
|x,z; j \rb = \sum_{\bf n, A} x^{n/2} \left(\frac{z}{x}\right)^{p/2} Q_{j}^{-1}(n,p; {\bf n, A}) | {\bf n, A};j \rb\,,
\ee
which satisfies (\ref{GJconds}).

\subsection{Pure $\SU(N)$}
The above construction also extends to the pure  $\SU(N)$ theory. The relevant class of descendants is
\be
|\vec{n},p;j \rb = (J_{-1}^{N-1,+})^p (J_0^{N-2,N-1 })^{n_{N-1}} \cdots   (J_0^{1,2})^{n_{2}} (J_{0}^{1,-})^{n_1} |j\rb \,.
\ee
We propose that the instanton expansion of the pure $\SU(N)$ theory in the presence of a full surface operator can be obtained from 
\be \label{propN}
 Z_{\rm inst}=\sum_{n_1=0}^{\infty} \cdots  \sum_{n_{N-1} =0}^{\infty} \sum_{p=0}^{\infty} Q_j^{-1}(\vec{n},p;\vec{n},p) \, x_1^{n_1} \left(\frac{x_2}{x_1}\right)^{n_2} \!\!\!  \cdots   \left(\frac{x_{N-1}}{x_{N-2 }}\right)^{n_{N-1}}\left(\frac{z}{x_{N-1}}\right)^p.
\ee
Again it is easy to check that the terms involving only one of the $N$ variables match with the instanton results. From  an earlier section we know that 
\bea
 \lb j| (J_0^{1+})^n  (J_0^{1-})^n | j \rb &=&n! (2\lb u_2-u_1,j\rb )_n(-1)^n \,, \non  \\[3pt]
 \lb j| (J_0^{i,i-1})^n  (J_0^{i-1,i})^n | j \rb &=&  n! (2\lb u_{i+1}-u_i,j\rb )_n(-1)^n \quad (i=2\ldots,N-1)\,, \non \\ 
 \lb j| (J_1^{N-1,-})^n  (J_{-1}^{N-1,+})^n | j \rb &=& n! (2 \lb u_1-u_N,j\rb   -k)_n(-1)^n \,.
\eea
Implementing these results into our proposal (\ref{propN}) and using 
\be 
y_1 = -\frac{x_1}{(\ep_1)^2} \,, \quad \; \;  y_{i+1}=-\frac{1}{(\ep_1)^2}\frac{x_{i+1}}{x_{i}} \quad (1\le i \le N{-}2) \,, \qquad y_{N}=-\frac{1}{(\ep_1)^2}\frac{z}{ x_{N-1}}\,,
\ee
together with the identifications 
\be
 \lb u_i , j\rb = \frac{1}{2}\lb u_i,\frac{a}{\ep_1}-\rho\rb  =  \frac{1}{2}(\,\frac{a_i}{\ep_1} - \half[N-2i+1])   \,, \qquad  \quad k=-N-\frac{\epsilon_2}{\epsilon_1} \,,
\ee
we find 
\be
 \sum_{n=0}^{\infty} \frac{1}{  n! \, ( a_{i+1}-a_i + \ep_1 + \ep_2 \lfloor \frac{i}{N} \rfloor   )_n } y_i^n \,,
\ee
which agrees with the instanton result (\ref{Z0pure}). Terms of the form $y_i^n y_j$ can also be matched, but we will not give the details here. 

Finally, the analogue of the Gaiotto state should satisfy
\bea \label{GJcondsN}
&&J_0^{1+} |\vec{x},z;j\rb = \sqrt{x_1} \,|\vec{x},z;j\rb \,, \qquad \qquad J_1^{N{-}1,-} |\vec{x},z;j\rb = \sqrt{\frac{z}{x_{N-1}}}\, |\vec{x},z;j\rb \,,  \\
&&J_0^{1,2} |\vec{x},z;j\rb =\sqrt{\frac{x_2}{x_1}}\, |\vec{x},z;j\rb \,, \quad \cdots \quad J_0^{N-2,N-1} |\vec{x},z;j\rb =\sqrt{\frac{x_{N-1}}{x_{N-2}}}\, |\vec{x},z;j\rb\,, \non
\eea
where all other $J_n^A$'s that annihilate $|j\rb$ also annihilate $|x,z;j\rb$, and has the expansion
\be 
|\vec{x},z; j \rb = \sum_{\bf n, A}  \, x_1^{n_1/2} \left(\frac{x_2}{x_1}\right)^{n_2/2} \!\!\!  \cdots   \left(\frac{x_{N-1}}{x_{N-2 }}\right)^{n_{N-1}/2}\!\! \left(\frac{z}{x_{N-1}}\right)^{p/2}  Q_{j}^{-1}(\vec{n},p; {\bf n, A}) | {\bf n, A};j \rb\,.
\ee
Let us finally mention that it  should be possible to derive the above construction as a limit of a conformal $\SU(N)$ theory when the masses are taken to infinity.  Note that in the above analysis the operator $\cK^{\dag}$ played no role, but it may be necessary for more general non-conformal quivers.

\setcounter{equation}{0} 
\section{Summary and outlook} \label{sdisc}

In this paper, building on earlier work \cite{Alday:2010}, we proposed a relation between instanton partition functions in $\SU(N)$ quiver gauge theories in the presence of a full surface operator (realised by a $4d$ defect from the $6d$ viewpoint ) and (slightly modified) affine $\sll(N)$ conformal blocks. Although this proposal passed several highly non-trivial tests it is still conjecture.  Possibly one can obtain a proof in special cases, e.g.~for the one-point conformal block on the torus along the lines in \cite{Fateev:2009} (extending the result in \cite{Negut:2008}). Perhaps the most important open problem is to gain a better understanding of the operator $\cK^\dag$.   

In the main text we did not specify precisely what the theory is whose conformal blocks reproduce the instanton partition function in the presence of a full surface operator. The reason is that the conformal blocks are completely determined by the symmetry algebra alone. Therefore knowledge of the precise theory was not needed.  However, just as in the AGT relation \cite{Alday:2009a}, one can speculate that the perturbative piece in the full partition function (involving some  extension of \cite{Pestun:2007}) may be related to the three-point parts of the correlation functions. 
Models with affine $\sll(2)$ symmetry include the $H_3^+$ (or $\SL(2,\CC)/\SU(2)$) WZNW model \cite{Teschner:1997}, as well as the $SL(2,\RR)$ WZNW model (see e.g.~\cite{Maldacena:2001}).  

In section \ref{m2m5} we checked that at the level of the instanton partition function
there is no distinction between surface operators arising from $2d$ and
$4d$ defects for theories with  gauge group $\SU(2)$.
However, the two realizations seem to be distinct  for quiver gauge
theories, already for  gauge theories  with two  $\SU(2)$ factors.
Nevertheless, it is known that affine $\sll(2)$ correlation functions and
Liouville correlation functions with degenerate field insertions are
related \cite{Ribault:2005}.  In this map the number of degenerate field
insertions ($2d$ defect surface operators) is larger than the number
expected for the description of a $4d$ defect surface operator, that
couple to all the gauge group factors in the quiver. However, there is a
modification of the map  \cite{Ribault:2005b} which requires fewer
degenerate field insertions provided one also modifies the affine
correlation functions. To obtain the right number of degenerate field
insertions ($2d$ surface operators) expected for a potential description of a $4d$ surface
operator,  one needs to replace one of the primary fields by its
spectrally-flowed \cite{Maldacena:2001} version with one unit of spectral
flow.
This could possibly be an alternative to the insertion of the operator
$\cK$ (in \cite{Minces:2005} some perturbative computations were performed
for the the four-point conformal block where one of the primary field is
spectrally flowed by one unit).
For the higher rank case it looks more problematic to relate affine
conformal blocks to  conformal blocks involving degenerate primaries in
Toda theories \cite{Ribault:2008}.  

In this paper we only discussed $4d$ $\SU(N)$ quiver gauge theories, but it should also be possible to study the corresponding $5d$ versions. The $5d$ instanton partition functions should be related to topological string partition functions. An important problem is to understand what a full surface operator arising from $4d$ defects  corresponds to in the topological string language (the topological string description of a simple surface operator was discussed in \cite{Kozcaz:2010,Dimofte:2010}; see also \cite{Taki:2010}). It would also be nice to find a matrix model description \cite{Dijkgraaf:2009}.

In the recent developments in $\cN=2$ gauge theories, the set of partitions of $N$ (or  
equivalently, the number of embeddings of $\SU(2)$ inside $\SU(N)$) appears in many places:  in the classification of punctures \cite{Gaiotto:2009a};  in the classification of the corresponding  degenerate states in the Toda field theories \cite{Kanno:2009}; and also in the classification of surface operators \cite{Gukov:2006}.  As we now recall, there is yet another place where the same classification appears. It is known that one can obtain the (quantum) $A_{N-1}$ Toda field theories from a WZNW model by so called Drinfeld-Sokolov reducton  (or hamiltonian reduction), see e.g.~\cite{Fateev:1987}. In this reduction, the affine $\sll(N)$ algebra turns into the $\cW_{N}$ algebra. What is perhaps less well known is that when the rank is larger than one there are in general many possible reductions. The various possibilities are classified by the number of embeddings of $\SU(2)$ inside $\SU(N)$ (see e.g.~\cite{Feher:1992}). The reduction that gives the standard Toda theory/$\cW_{N}$ algebra \cite{Fateev:1987} corresponds to the principal embedding. The simplest example not of this form arises for rank two and leads to the Polyakov-Bershadsky algebra $\cW_3^{(2)}$ \cite{Polyakov:1989}. One may wonder if chiral 
 blocks in these more general $\cW$ algebras have a dual gauge theory interpretation.

\section*{Acknowledgements}

The research of FP was supported by the Volkswagen Foundation.
NW would like to thank the string theory group at IST, Lisbon, where part of this work was done, for generous hospitality and financial support. He would also like to thank the organisers for invitations to the following inspiring conferences: `16 Supersymmetries', London;  `17th Irish Quantum Field Theory Meeting', Dublin;  `Integrability in Gauge and String Theory 2010', Stockholm.


\appendix

\setcounter{equation}{0} 
\section{Appendix}

\subsection{The $\sll(N)$ Lie algebras} \label{ALie}

Here we summarise some standard results for the $\sll(N)$ (or $A_{N-1}$) Lie algebras.
The root/weight space of the $\sll(N)$ Lie algebra can viewed as a $N-1$--dimensional subspace of $\RR^{N}$. The unit vectors of $\RR^{N}$ will be denoted $u_i$ ($i=1,\ldots,N$) and satisfy $\lb u_i,u_j\rb = \de_{ij}$. The simple roots are $e_i = u_i-u_{i+1}$ ($i=1,\ldots,N-1$) and the positive roots are $e_{ij} =  u_i-u_j$ (with $1\leq i<j \leq {N}$). 
The Weyl vector, $\rho$, is half the sum of the positive roots; hence $\rho = \half \sum_{i=1}^{N} (N- 2i + 1) u_i$. The fundamental weights, $\La_i$, are defined as
\be
\La_i = u_1+\cdots+u_i - \frac{i}{N}\sum_{j=1}^{N} u_j \,, \qquad (i=1,\ldots,N-1)
\ee
and  satisfy $\lb \La_i ,e_j\rb = \de_{ij}$. Note that $\sum_{i=1}^{N-1} \La_i = \rho$. Finally, the weights of the fundamental representation can be chosen as 
\be \label{hs}
h_i = u_i - \frac{1}{N} \sum_j u_j = \La_1 - \sum_{j=1}^{i-1} e_j\,, \qquad (i=1,\ldots,N)
\ee
Note that $h_1 = \La_1$ and $\sum_j h_j =0$. 

\subsection{Subleading terms in $Z_{\rm inst}$ for $\SU(N)$ with a full surface operator}\label{instsub}

Using the expressions given in section \ref{sinst}, we find that for the pure $\SU(N)$ theory the $y_i^n y_{j}$ terms in the instanton partition function are given by
\bea \label{Z1pure}
Z_{\rm inst}^{(1)i,j} &=& 
\sum_{n=1}^{\infty} \frac{ 1}{ (\frac{a_{i+1}}{\ep_1}-\frac{a_{i}}{\ep_1}+\frac{\ep_2}{\ep_1} \lfloor \frac{i}{N} \rfloor +1)_n \,  n! }
\bigg\{  \bigg[ \frac{  \frac{a_{i+1}}{\ep_1}-\frac{a_{i}}{\ep_1}+\frac{\ep_2}{\ep_1} \lfloor \frac{i}{N} \rfloor + n  }{  \frac{a_{i+1}}{\ep_1}-\frac{a_{i}}{\ep_1}+\frac{\ep_2}{\ep_1} \lfloor \frac{i}{N} \rfloor }         \bigg]^{\de_{j,i+1}} \non \\   
&& \quad \times 
\bigg[ \frac{  \frac{a_{j+1}}{\ep_1}-\frac{a_{j}}{\ep_1}+\frac{\ep_2}{\ep_1} \lfloor \frac{j}{N} \rfloor +1 }{   \frac{a_{j+1}}{\ep_1}-\frac{a_{j}}{\ep_1}+\frac{\ep_2}{\ep_1} \lfloor \frac{j}{N} \rfloor -n+1}  \bigg]^{\de_{i,j+1}}
 \frac{ 1 }{ \frac{a_{j+1}}{\ep_1}-\frac{a_{j}}{\ep_1}+\frac{\ep_2}{\ep_1} \lfloor \frac{j}{N} \rfloor +1 }  
\non \\
&+& \!\! {\de_{i,j+1}} \bigg[  \frac{ n\, ( \frac{a_{i+1}}{\ep_1}-\frac{a_{i}}{\ep_1}+\frac{\ep_2}{\ep_1} \lfloor \frac{i}{N} \rfloor + n )  }{  (\frac{a_{i+1}}{\ep_1}{-}\frac{a_{i-1}}{\ep_1}{+}\frac{\ep_2}{\ep_1} \lfloor \frac{j+1}{N} \rfloor {+}1) (\frac{a_{i}}{\ep_1}{-}\frac{a_{i-1}}{\ep_1}{+}\frac{\ep_2}{\ep_1} \lfloor \frac{j}{N} \rfloor {+}1) (\frac{a_{i-1}}{\ep_1}{-}\frac{a_{i}}{\ep_1}{-}\frac{\ep_2}{\ep_1} \lfloor \frac{j}{N} \rfloor {+}n{-}1)   } \bigg]
 \non \\
 &+&  \!\!  {\de_{j,i+1}} \bigg[ \frac{ n  }{ (\frac{a_{i+2}}{\ep_1}-\frac{a_{i}}{\ep_1}+\frac{\ep_2}{\ep_1} \lfloor \frac{i+1}{N} \rfloor +1) (\frac{a_{i}}{\ep_1}-\frac{a_{i+1}}{\ep_1}-\frac{\ep_2}{\ep_1} \lfloor \frac{i}{N} \rfloor )  }
  \bigg] \bigg\} 
  \left(\frac{y_i}{(\ep_1)^2}\right)^n \frac{y_j}{(\ep_1)^2} \, .
\eea
In the above expression, the Kronecker $\de$ is periodically definied i.e.~$\de_{i,j}=\de_{i+N,j}=\de_{i,j+N}$. Similarly, for the conformal  $\SU(N)$ theory with $N_f=2N$ we find
\bea \label{nf41}
Z_{\rm inst}^{(1)i,j} &=& 
\sum_{n=1}^{\infty} \frac{ (\frac{\mu_{i+1}}{\ep_1}-\frac{a_{i}}{\ep_1}+\frac{\ep_2}{\ep_1} \lfloor \frac{i}{N} \rfloor +1 )_n(\frac{\tilde{\mu}_i}{\ep_1}-\frac{a_{i}}{\ep_1})_n }{ (\frac{a_{i+1}}{\ep_1}-\frac{a_{i}}{\ep_1}+\frac{\ep_2}{\ep_1} \lfloor \frac{i}{N} \rfloor +1)_n \,  n! }
\bigg\{  \bigg[ \frac{  \frac{a_{i+1}}{\ep_1}-\frac{a_{i}}{\ep_1}+\frac{\ep_2}{\ep_1} \lfloor \frac{i}{N} \rfloor + n  }{  \frac{a_{i+1}}{\ep_1}-\frac{a_{i}}{\ep_1}+\frac{\ep_2}{\ep_1} \lfloor \frac{i}{N} \rfloor }         \bigg]^{\de_{j,i+1}} \non \\   
&& \; \times 
\bigg[ \frac{  \frac{a_{j+1}}{\ep_1}-\frac{a_{j}}{\ep_1}+\frac{\ep_2}{\ep_1} \lfloor \frac{j}{N} \rfloor +1 }{   \frac{a_{j+1}}{\ep_1}-\frac{a_{j}}{\ep_1}+\frac{\ep_2}{\ep_1} \lfloor \frac{j}{N} \rfloor -n+1}  \bigg]^{\de_{i,j+1}}
 \frac{  (\frac{\mu_{j+1}}{\ep_1}-\frac{a_{j}}{\ep_1}+\frac{\ep_2}{\ep_1} \lfloor \frac{j}{N} \rfloor +1 ) ( \frac{\tilde{\mu}_{j}}{\ep_1} - \frac{a_{j}}{\ep_1} ) }{ \frac{a_{j+1}}{\ep_1}-\frac{a_{j}}{\ep_1}+\frac{\ep_2}{\ep_1} \lfloor \frac{j}{N} \rfloor +1 }  
\non \\
 && 
\!\!\!\!\!\!\!\!\!\!\! + \,{\de_{i,j+1}}  \bigg[  
\frac{ n\, 
( \frac{a_{i+1}}{\ep_1}{-}\frac{a_{i}}{\ep_1}{+}\frac{\ep_2}{\ep_1} \lfloor \frac{i}{N} \rfloor {+} n )  }{  
(\frac{a_{i+1}}{\ep_1}{-}\frac{a_{i-1}}{\ep_1}{+}\frac{\ep_2}{\ep_1} \lfloor \frac{j+1}{N} \rfloor {+}1) (\frac{a_{i}}{\ep_1}{-}\frac{a_{i-1}}{\ep_1}{+}\frac{\ep_2}{\ep_1} \lfloor \frac{j}{N} \rfloor {+}1)
 (\frac{a_{i-1}}{\ep_1}{-}\frac{a_{i}}{\ep_1}{-}\frac{\ep_2}{\ep_1} \lfloor \frac{j}{N} \rfloor {+}n{-}1)   } \bigg]
 \non \\
 && \!\!\!\!\!\!\!\!\!\!\! \times \bigg[  
\frac{ n\, 
(\frac{\mu_{i+1}}{\ep_1}{-}\frac{a_{i-1}}{\ep_1}{+}\frac{\ep_2}{\ep_1} \lfloor \frac{j{+}1}{N} \rfloor {+}1)  (\frac{\mu_{i}}{\ep_1}{-}\frac{a_{i-1}}{\ep_1}{+}\frac{\ep_2}{\ep_1} \lfloor \frac{j}{N} \rfloor {+}1)        
 (\frac{\tilde{\mu}_{i-1}}{\ep_1}{-}\frac{a_{i-1}}{\ep_1}) 
 (\frac{\tilde{\mu}_{i}}{\ep_1}{-}\frac{a_{i-1}}{\ep_1}{+}\frac{\ep_2}{\ep_1} \lfloor \frac{j}{N} \rfloor ) 
 }{  
(\frac{\mu_{i+1}}{\ep_1}{-}\frac{a_{i}}{\ep_1}{+}\frac{\ep_2}{\ep_1} \lfloor \frac{i}{N} \rfloor {+}n) 
(\frac{\tilde{\mu}_{i}}{\ep_1}{-}\frac{a_{i}}{\ep_1} {-}n{+}1) 
  } \bigg]
 \non \\
 &+&  \!\! {\de_{j,i+1}}   \bigg[ \frac{ n\,  (\frac{\mu_{i+2}}{\ep_1}{-}\frac{a_{i}}{\ep_1}{+}\frac{\ep_2}{\ep_1} \lfloor \frac{i+1}{N} \rfloor {+}1 ) ( \frac{\tilde{\mu}_{i+1}}{\ep_1} {-} \frac{a_{i}}{\ep_1}{+} \frac{\ep_2}{\ep_1} \lfloor \frac{i}{N} \rfloor )  }{ (\frac{a_{i+2}}{\ep_1}-\frac{a_{i}}{\ep_1}+\frac{\ep_2}{\ep_1} \lfloor \frac{i+1}{N} \rfloor +1) (\frac{a_{i}}{\ep_1}-\frac{a_{i+1}}{\ep_1}-\frac{\ep_2}{\ep_1} \lfloor \frac{i}{N} \rfloor )  }
  \bigg]\bigg\} 
  (-y_i)^n \, (-y_j) \,.
\eea

\subsection{Computations of affine conformal blocks: technical details}\label{appmix}

In this appendix we collect some selected details of the computations of affine conformal blocks performed in sections \ref{ssl2} and \ref{sslN}. 

A rearrangement formula that we repeatedly used is the Zassenhaus formula
\be \label{Zassy}
 e^{X+Y} = e^X e^Y  e^{-\frac{1}{2}[X,Y]} e^{\frac{1}{6}(2[Y,[X,Y]]+[X,[X,Y]])} \mathellipsis
\ee
In the computations we also repeatedly used manipulations of the type  
\be
 (J_0^+)^p (J_0^-)^n   |j \rb  =  n\,(2j-n+1) \,(J_0^+)^{p-1}(J_0^-)^{n-1}   |j \rb \,.
\ee
As an example, for the $x^n$ terms in the $\widehat{\sll}(2)$ four-point conformal block on the sphere the piece involving $\cK$ is computed as follows:
\bea
&& \!\!\!\!\!\!\!\!\! \lb j | (J_0^+)^n e^{ -x J_0^-}  V_{j_3}(x,z) |j_4 \rb   
 = \sum_{p=0}^n \frac{(-x)^p}{p!} \lb j | (J_0^+)^n ( J_0^-)^p  V_{j_3}(x,z) |j_4 \rb \non \\
 &=& \sum_{p=0}^n \frac{(-x)^p}{p!}   \frac{ n! \, (-2j)_n \, (-1)^n }{(n-p)! \, (-2j)_{n-p} \, (-1)^{n-p}  } \lb j | (J_0^+)^{n-p}  V_{j_3}(x,z) |j_4 \rb  \non \\
 &=&  n! \,(-2j)_n   \sum_{p=0}^n \frac{(-x)^p}{p!}  \frac{  (-1)^p }{(n-p)! (-2j)_{n-p}  } (j_4-j_3-j)_{n-p} (-x)^{n-p}    \\
 &=&   n! \,(-2j)_n \, (-x)^n \,   \sum_{p=0}^n \frac{ (-1)^p }{p!}  \frac{ (j_4-j_3-j)_{n-p} }{(n-p)! (-2j)_{n-p}  } 
 =  x^n \, (-2j-[j_4-j_3-j] )_{n}  \,. \non 
 \eea
For the mixed term of the form $z \, x^n$ in the  $\widehat{\sll}(2)$  four-point conformal block on the sphere the terms one needs to compute are:
\bea
\lb 1|\cK(x,z) V_{j_3}(x,z) |j_4 \rb &=&z \, x^{n} \, (k-j_3-j_4-j+n+1) (j_3-j_4-j)_{n+1}\, , \non \\
\lb 2| \cK(x,z)  V_{j_3}(x,z) |j_4 \rb &=&z\, x^n \, (j+j_4-k/2-n) (j_3-j_4-j)_n\, , \\
\lb 3|   \cK(x,z) V_{j_3}(x,z) |j_4 \rb  &=& z\, x^n \, (-j_3+j_4+j-n+1) (j_3-j_4-j)_{n-1}  \,, \non
\eea
as well as
\bea
\lb j_1| V_{j_2}(1,1)  |1  \rb&=&(j-j_2-j_1-n-1)(j_1-j_2-j)_{n+1}\, , \non \\
 \lb j_1| V_{j_2}(1,1)   |2 \rb&=&(j-j_1-n) (j_1-j_2-j)_n \, , \\
 \lb j_1| V_{j_2}(1,1)  |3  \rb&=&(j_1-j_2-j+n-1)(j_1-j_2-j)_{n-1} \, . \non
\eea
Finally, the  $3{\times}3$ Gram matrix $X_{rs} = \lb r | s \rb$ becomes:
\be \label{G33}
\left( \ba{ccc} [k{-}2j{+}2n{+}2] M(n+1) & M(n{+}1)& 0 \\M(n{+}1) &\frac{k}{2} M(n) & -M(n) \\ 0 & -M(n) & [k{+}2j{-}2n{+}2)] M(n{-}1) \ea \right)
\ee
where 
\be
M(n)\equiv {\lb j |  (J_0^{-} )^{n}     (J_0^{-} )^{n}   | j \rb }=(-2j)_n n! (-1)^n\,.
\ee
The computation of the $x_1^n x_2$ terms in the $\widehat{\sll}(N)$ four-point conformal block is very similar. The relevant descendants are
\be
|1\rb=J^{12}_0 (J_0^{1-})^n   | j \rb, \qquad  |2\rb= J_0^{2-} (J_0^{1-})^{n-1}    | j \rb\,,
\ee
and the terms one needs to compute are:
\bea
\nonumber \lb j_1| V_{\chi_2}(1,1)\mathcal{K}^{\dagger} (1,1) | 1 \rb&\!\!=\!\!&  (-1)^n    \left( {\ts \frac{2\kappa_2}{N}} {+}2\lb h_2 , j_1\rb {-} 2 \lb h_1, j \rb\right)_n \! \left({\ts \frac{2\kappa_2}{N} } {-}2\lb h_2 , j\rb {+}2 \lb h_3, j_1 \rb{-}n\right) \\
\lb j_1| V_{\chi_2}(1,1)\mathcal{K}^{\dagger} (1,1) | 2 \rb&\!\!=\!\!&  (-1)^n     \left( {\ts \frac{2\kappa_2}{N}} +2\lb h_2 , j_1\rb - 2 \lb h_1, j \rb\right)_n \, ,
\eea
as well as
\bea
\nonumber
 \lb 1 | V_{\chi_3}(x,z)     | j_4 \rb&\!\!=\!\!&  ({-}x_1)^{n-1} x_2     \left(-{\ts \frac{2\kappa_3}{N}}{+}2\lb h_1 , j_4\rb {-} 2 \lb h_1, j \rb\right)_n  \left({\ts \frac{2\kappa_3}{N}}{-}2\lb h_2 , j_4\rb {+} 2 \lb h_2, j \rb{+}n\right) \\
 \lb 2 | V_{\chi_3}(x,z)     | j_4 \rb&\!\!=\!\!& ({-}x_1)^{n} x_2  \left(-{\ts \frac{2\kappa_3}{N}}{+}2\lb h_1 , j_4\rb {-} 2 \lb h_1, j \rb\right)_n  \, .
\eea
Finally, the  $2{\times}2$ Gram matrix $X_{rs} = \lb r | s \rb$ with $r,s=1,2$ becomes:
\be \label{G22}
\left( \ba{cc}    \left(2\lb h_2 , j\rb {-} 2 \lb h_3, j \rb{+}n\right)  S(n) & - S(n) \\ - S(n)  &   \left(2\lb h_1 , j\rb {-} 2 \lb h_3, j \rb{-}n{+}1 \right)  S(n-1) \ea \right)
\ee
where 
\be
S(n)\equiv  (-1)^n n!         \left(-2\lb h_1 , j\rb + 2 \lb h_2, j \rb\right)_n  \,.
\ee

\subsection{Liouville conformal blocks  with degenerate  operators}\label{liouville}
Consider a five-point Liouville conformal  block  where  one  of  the  insertions  is  a  degenerate field, i.e. 
\be \label{cbliou}
\lb \al_1| V_{\al_2}(1)  V_{-\frac{b}{2}}(x) V_{\al_3}(z) |\al_4 \rb\, . 
\ee
We  insert  two complete  sets  of  states,  yielding    
\bea \label{liouex}
\!\! \sum_{{\bf n},{\bf n}',{\bf p},{\bf p}'} \!\! \lb \al_1| V_{\al_2}(1) |{\bf n};\si \rb X^{-1}_{ \bf n; \bf n'}(\si)
\lb {\bf n}';\si|V_{-\frac{b}{2} }(x) |{\bf p};\tilde{\si} \rb X^{-1}_{ \bf p; \bf p'}(\tilde{\si})
\lb {\bf p}';\tilde{\si} | V_{\al_3}(z) |\al_4 \rb \,, 
\eea 
where the  sum  is over  partitions ${\bf n}=(n_1,n_2,\ldots )$ with $1\leq n_{1}\leq
n_{2}\leq\mathellipsis\leq n_{r}$ and  $|{\bf n}; \si\rb$ are descendants of the primary
state  $|\si\rb$,  i.e.~$ |\si,{\bf n}\,\rangle\equiv L_{-n_{1}}L_{-n_{2}}\mathellipsis
L_{-n_{r}}|\si\rangle$.   $X^{-1}_{ \bf n; \bf n'}(\si)$  is  the  inverse of  the Gram matrix  $X_{ \bf n; \bf n'}(\si)=    \lb \bf n; \si |\bf n' ; \si \rb $.
The  matrix $X_{ \bf p; \bf p'}(\tilde{\si})$
and the states  $|{\bf p}; \tilde{\si} \rb$ are defined  in a similar  way.   The  terms in (\ref{liouex}) with ${\bf p}={\bf p'}=0$ depend only on  $x$ and sum  up  to  $\lb \al_1| V_{\al_2}(1) V_{-\frac{b}{2} }(x) |\tilde{\si} \rb$.  The  BPZ \cite{Belavin:1984}  equation implies that 
\bea \label{mzero}
 \lb \al_1| V_{\al_2}(1) V_{-\frac{b}{2} }(x) |\tilde{\si}\rb=x^{b\tilde{\si}}(1-x)^{b\al_2}  \, G(x)
  \eea
where $G(x)$ satisfies the hypergeometric   differential equation. The solution defined  in  a neighbourhood  of  $x=0$ that we need is 
\be \label{gz}
G(x)= \,{}_2 F_1(A,B;C;x)
\ee
with 
\be \label{ABC2}
A = b\, (-\al_1+\al_2+\tilde{\si}-\frac{b}{2})\,,\quad  B=b\, (\al_1+\al_2+\tilde{\si} - \cQ -\frac{b}{2} )\,, \quad C=b\, ( 2 \tilde{\si} - b) \, ,
\ee
where $\cQ=b+\frac{1}{b}$.  In  order  to  match  this  component  of  the  conformal block to the  instanton  partition  function with $y_2=0$, cf.~(\ref{Nf4Z0}),  
we use  the  relations $\epsilon_1=\frac{1}{b}$, $\epsilon_2=b$ and 
\bea
x=-y_1,\quad\alpha_1=\frac{b}{2}+\frac{\tilde{\mu}_1-\mu_2}{2},\quad\alpha_2=\frac{\cQ}{2}+\frac{\tilde{\mu}_1+\mu_2}{2},\quad\tilde{\si}=\frac{\cQ}{2}-a_1\,.
\eea 
  
The  terms in (\ref{liouex}) with ${\bf n}={\bf n'}=0$ are a power series in $\frac{x}{z}$ that sums  up  to $\lb \si | V_{-\frac{b}{2} }(x) V_{\al_3}(z) |\al_4 \rb $. Imposing the  BPZ equation  we  have 
\bea \label{mzeroy}
\lb \si | V_{-\frac{b}{2} }(x) V_{\al_3}(z) |\al_4 \rb =\left(\frac{x}{z}\right)^{b\alpha_4}\left(1-\frac{x}{z}\right)^{b\al_3}  \, H\left(\frac{x}{z}\right)
  \eea
where $H\left(\frac{x}{z}\right)$ satisfies the hypergeometric   differential equation. 
We  consider  the solution  that  is  defined  around  $\frac{x}{z}=\infty$; in  details
\be \label{hy}
H\left(\frac{x}{z}\right)= \left(\frac{x}{z}\right)^{-C_1}\,{}_2 F_1(C_1,C_1+1-D_1;C_1-C_2+1;\frac{z}{x}) \,,
\ee
where
\be \label{cd}
C_1 = b\, (-\sigma+\al_3+\alpha_4-\frac{b}{2})\,,\quad  C_2=b\, (\sigma+\al_3+\alpha_4 - \cQ -\frac{b}{2} )\,, \quad D_1=b\, ( 2 \alpha_4 - b) \,.
\ee
Considering  the  dictionary 
\bea
\frac{z}{x}=-y_2,\quad\alpha_3=\frac{\cQ}{2}+\frac{\mu_1+\tilde{\mu}_2}{2},\quad\alpha_4=\cQ+\frac{\mu_1-\tilde{\mu}_2}{2},\quad\sigma=\frac{\cQ}{2}-a_1-\frac{b}{2}
\eea 
we  reproduce  the instanton partition  function  (\ref{Nf4Z0}) depending on  $y_2$.

We have also computed the $z\,x^m$ terms with the result that  the instanton partition function $Z$  is  equal  to the  Liouville block  up  to  a prefactor, i.e.
\bea
Z& =&(1-z)^{W}(1-\frac{z}{x})^{2 b \alpha_3}
 x^{b\tilde{\si}}(1-x)^{b\alpha_2}  \lb \alpha_1| V_{\alpha_2}(1)  V_{b/2}(x)V_{\alpha_3}(z) |\alpha_4 \rb
\,,
\eea
where 
\bea\nonumber
W=-\al_2\al_3+\frac{3}{8}(2\al_1-2\al_2-2\al_3-2\al_4+b)(-2\al_1-2\al_2-2\al_3+2\al_4+b)\,.
\eea
This  result was obtained by expressing the Virasoro $L_{-n_i}$ operators  as differential  operators and   showing   that 
\bea
Z& =&(1-z)^{W}       (1-\frac{z}{x})^{2 b \alpha_3}(G(x)+ z~ T^{(1)}_x+ z^2~ T^{(2)}_x +\cdots)\eea
where $G(x)$ is the hypergeometric function defined in (\ref{gz}) and
\be
T^{(1)}_x=\frac{1}{x^{b\tilde{\si}}(1-x)^{b\alpha_2} }\nabla_x\left(   x^{b\tilde{\si}}(1-x)^{b\alpha_2} G(x)              \right),
\ee
with
\be
\nabla_x\equiv \left[(\Delta(\tilde{\si}){+}\Delta(\alpha_3) {-}\Delta(\alpha_4)\right]\frac{1}{\Delta(\tilde{\si})}
[  (1-x)\partial_x{+}\Delta(\alpha_1){-}\Delta(\alpha_2){-}\Delta(\tilde{\si}){-}\Delta(-\frac{b}{2})  ] \,.
\ee

\begingroup\raggedright\endgroup

\end{document}